\documentclass{aa}  
\usepackage{txfonts}
\usepackage{graphicx}   
\usepackage{tabularx}
\usepackage{subfigure}
\usepackage[colorlinks=true,allcolors=blue]{hyperref}

\begin{document} 

\newcommand{\ha}{\hbox{H$\alpha$}}
\newcommand{\hb}{\hbox{H$\beta$}}
\newcommand{\hg}{\hbox{H$\gamma$}}
\newcommand{\hd}{\hbox{H$\delta$}}
\newcommand{\he}{\hbox{H$\varepsilon$}}
\newcommand{\hz}{\hbox{H$\zeta$}}
\newcommand{\hei}{\hbox{He\,{\sc i}}}
\newcommand{\heii}{\hbox{He\,{\sc ii}}}
\newcommand{\oi}{\hbox{[O\,{\sc i}]}}
\newcommand{\oii}{\hbox{[O\,{\sc ii}]}}
\newcommand{\oiii}{\hbox{[O\,{\sc iii}]}}
\newcommand{\nii}{\hbox{[N\,{\sc ii}]}}
\newcommand{\sii}{\hbox{[S\,{\sc ii}]}}
\newcommand{\siii}{\hbox{[S\,{\sc iii}]}}
\newcommand{\feii}{\hbox{[Fe\,{\sc ii}]}}
\newcommand{\neiii}{\hbox{[Ne\,{\sc iii}]}}
\newcommand{\niii}{\hbox{[Ni\,{\sc ii}]}}
\newcommand{\caii}{\hbox{[Ca\,{\sc ii}]}}

   \title{Ionized gas in quiescent galaxies: Temperature measurement and constraint on the ionization source}

   \subtitle{}

   \author{Man-Yin Leo Lee \inst{1}\thanks{Email: myleolee@link.cuhk.edu.hk},
          Renbin Yan \inst{1}\thanks{Email: rbyan@phy.cuhk.edu.hk},
          Xihan Ji \inst{2, 3, 4},
          Gerome Algodon \inst{5},
          Kyle Westfall \inst{5},
          Zesen Lin \inst{1},
          Francesco Belfiore \inst{6}
          \and
          Kevin Bundy\inst{5}
          }

   \institute{Department of Physics, The Chinese University of Hong Kong, Shatin, New Territories, Hong Kong SAR, China
        \and
            Kavli Institute for Cosmology, University of Cambridge, Madingley Road, Cambridge, CB3 0HA, UK
        \and
            Cavendish Laboratory, University of Cambridge, 19 JJ Thomson Avenue, Cambridge, CB3 0HE, UK
        \and
            Department of Physics and Astronomy, University of Kentucky, 505 Rose Street, Lexington, KY 40506, USA
        \and
            UC Observatories, MS: UCO Lick, UC Santa Cruz, 1156 High St, Santa Cruz, CA 95064, USA
        \and
            INAF - Osservatorio Astrofisico di Arcetri, Largo E. Fermi 5, I50125, Florence, Italy
             }

   \date{Received November 1, 2023; accepted July 1, 2024}

 
  \abstract{In non-star-forming, passively evolving galaxies, regions with emission lines dominated by low-ionization species are classified as low-ionization emission regions (LIERs). The ionization mechanism behind such regions has long been a mystery. Active galactic nuclei (AGNs), which were once believed to be the source, have been found not to be the dominant mechanism, especially in regions distant from the galaxy nuclei. The remaining candidates, photoionization by post-asymptotic giant branch (pAGB) stars and interstellar shocks can only be distinguished with in-depth analysis. As the temperature predictions of these two models differ, temperature measurements can provide strong constraints on this puzzle. We selected a sample of 2795 quiescent red-sequence galaxies from the Sloan Digital Sky Survey IV (SDSS-IV) Mapping Nearby Galaxies at Apache Point Observatory (MaNGA) survey. We divided the sample spectra into three groups based on the \nii/\ha\ flux ratio, and utilized stacking techniques to improve the signal-to-noise ratio of the observed spectra. We determined the temperature of \oiii, \nii, \sii, and \oii\ through their temperature-sensitive emission line ratios. Subsequently, we compared the measured temperatures with predictions from different models. The results demonstrate consistency with the interstellar shock model with  pre-shock density n = 1 cm$^{-3}$ and solar metallicity, thus supporting shocks as the dominant ionization source of LIERs. Additionally, we also find that the interstellar dust extinction value measured through the Balmer decrement appears to be larger than that implied by the forbidden line ratios of low-ionization lines.}


   \titlerunning{Temperature measurements of LIERs}
   \authorrunning{Lee et al.}
   \maketitle
%

\section{Introduction}
Non-star-forming, passively evolving galaxies are often referred to as quiescent galaxies (QGs). These galaxies are characterized by their red optical  colors and are dominated by evolved stellar populations. Originally, QGs were thought to be poor in cool and warm gas. Nevertheless, due to improvements in observations, it is found that diffuse warm ionized gas is present in such galaxies \citep{knappStatisticalDistributionNeutralhydrogen1985, phillipsIonizedGasElliptical1986, kimInterstellarMatterEarlyType1989, busonTgheDistributionIonized1993, goudfrooijInterstellarMatterShapleyAmes1994, knappMolecularGasElliptical1996, macchettoSurveyISMEarlytype1996, zeilingerDistributionIonizedGas1996, sarziSAURONProjectIntegralfield2006, davisATLAS3DProjectOrigin2011, kehrigIonizedGasCALIFA2012}. As revealed by \cite{phillipsIonizedGasElliptical1986}, the spectrum of such gas exhibits strong emission lines from low-ionization ions, similar to low-ionization nuclear emission-line regions (LINERs) \citep{heckmanOpticalRadioSurvey1980}. Ionization of LINERs was widely believed to be related to the central active galactic nuclei (AGNs) in the galaxy \citep{ferlandAreThereAny1983, halpernLowIonizationActive1983, hoSupermassiveBlackHoles1999, kauffmannHostGalaxiesActive2003, grovesDustyRadiationPressureDominated2004, kewleyHostGalaxiesClassification2006, hoOriginDynamicalSupport2009} and the ionized interstellar medium (ISM) in QGs were therefore thought to have the same ionization mechanism.

In recent years, observations with enhanced spatial and spectral resolution have led to discoveries of similar emission regions extending outside the nuclear region of galaxies \citep{sarziSAURONProjectXVI2010, yanNATURELINERLIKEEMISSION2012, singhNatureLINERGalaxies2013, belfioreSDSSIVMaNGA2016}, raising a request for a different ionization mechanism. Following \cite{belfioreSDSSIVMaNGA2016}, these regions are referred to as low-ionization emission regions (LIERs). While the dominant ionizing source remains a mystery, two main candidates have been proposed, which include photoionization by post-asymptotic giant branch (pAGB) stars \citep{binettePhotoionizationEllipticalGalaxies1994, stasinskaCanRetiredGalaxies2008, sarziSAURONProjectXVI2010, cidfernandesComprehensiveClassificationGalaxies2011, yanNATURELINERLIKEEMISSION2012, singhNatureLINERGalaxies2013, belfioreSDSSIVMaNGA2016, yanShocksPhotoionizationDirect2018, bylerSelfconsistentPredictionsLIERlike2019} and collisional ionization by fast shocks \citep{heckmanOpticalRadioSurvey1980, dopitaSpectralSignaturesFast1995, grovesDustyRadiationPressureDominated2004, allenMAPPINGSIIILibrary2008}. 

The spectral emission line ratio is one of the tools that has been used to investigate the physical condition of the ISM. One of the most typical examples is the Baldwin-Phillips-Terlevich (BPT) diagram \citep{baldwinClassificationParametersEmissionline1981, veilleuxSpectralClassificationEmissionLine1987}. With different line ratios plotted against each other, we can characterize galaxies by their ionization mechanisms. Nonetheless, shocks and post-AGB stars can produce similar ratios in emission lines used by the BPT diagrams. Models of the two mechanisms cover similar areas on these diagrams, and thus do not provide any leverage to distinguish the dominant ionization source. In addition to the BPT diagrams, different emission line ratios can be used to measure the electron density, metallicity, ionization parameters, or temperature of specific ion species. The physical conditions of the ionized ISM are expected to depend on its ionization sources. Given that collisional ionization by interstellar shocks and photoionization by pAGB stars predict different temperatures in the gas, measuring the electron temperatures provides a way to distinguish between the two mechanisms. In general, gas ionized by interstellar shocks is expected to have a higher temperature than photoionized gas. The temperature of different ions can be measured by the ratio of auroral lines to strong lines from the gas.

For this study we made use of spectroscopic data from the Sloan Digital Sky Survey IV (SDSS-IV) Mapping Nearby Galaxies at Apache Point Observatory (MaNGA) survey to detect temperature-sensitive line ratios:  \oiii\ $\lambda$4363/\oiii\ $\lambda$5007, \nii\ $\lambda$5755/\nii\ $\lambda$6583, \sii\ $\lambda$4068, 4076/\sii\ $\lambda$6716, 6731,  and \oii\ $\lambda$7320, 7330/\oii\ $\lambda$3727, 3729. Following  procedures similar  to those used in \cite{yanShocksPhotoionizationDirect2018}, we made use of stacking techniques to increase the S/N of the observed spectrum. Because LIERs have weak emission lines in general and the temperature-sensitive auroral lines are even weaker, we needed very accurate subtraction of the stellar continua, in addition to high S/N, to detect these weak lines. Full-spectrum fitting using stellar population models turned out to be insufficient. Empirical spectra from the same dataset  produced a much better result. Thus, we selected those quiescent spaxels without emission lines, matching them to those spaxels with emission lines in terms of stellar population. From these line-free spaxels, we stacked their spectra together to get modeled stellar continua.  After subtracting the stellar continua, the fluxes of the emission lines were measured and the temperature-sensitive line ratios were compared to the prediction of models. 

This paper is organized as follows. In Sect. \ref{sec:data} we briefly describe the data. In Sect. \ref{sec: method} we present the methodology of our spaxel selection, stacking, and line flux measurement. In Sect. \ref{sec: data analysis} different physical properties of the data are computed. The results are then compared to the models assuming different ionization sources in Sect. \ref{sec: Discussion}. In Sect. \ref{sec: conclusions} we summarize our main findings. 

All absolute magnitudes in this paper are calculated with H$_0$ = 100$h$ km s$^{-1}$ Mpc$^{-1}$, so they should be interpreted as M - 5 $\log_{10}h$, in the AB magnitude system.\\
\section{Data}
\label{sec:data}
We used data from MaNGA survey \citep{bundyOVERVIEWSDSSIVMaNGA2014, yanSDSSIVMaNGAIFS2016}, which is part of the SDSS-IV \citep{blantonSloanDigitalSky2017}. The MaNGA instrument contains 17 integral field units (IFUs) which are composed of 19 to 127 fibers \citep{droryMaNGAIntegralField2015}. These IFUs cover diameters between 12 and 32 arcsecs in the sky. Using the 2.5m Sloan telescope \citep{gunnTelescopeSloanDigital2006} at the Apache Point Observatory, the light from the galaxies is guided through the fibers and fed into the BOSS spectrograph \citep{smeeMULTIOBJECTFIBERFEDSPECTROGRAPHS2013}, where they are split into blue and red channels. In each channel, the light is dispersed producing spectra with a resolution of R$\sim$2000 and a wavelength range of 3622 - 10354\AA. 

The SDSS Data Release 17 \citep{abdurroufSeventeenthDataRelease2022} contains the complete release of the MaNGA survey, consisting of over 10,000 nearby galaxies. Most of the targets are covered with radial coverage between 1.5 to 2.5 effective radii (R$_e$) \citep{wakeSDSSIVMaNGASample2017}. Since all these galaxies have a redshift of 0.01 < $z$ < 0.15, the wavelength coverage of MaNGA covers all the emission lines of interest in our project. The spectroscopic data we used is processed by the MaNGA Data Reduction Pipeline (DRP) \citep{lawDataReductionPipeline2016, lawSDSSIVMaNGAModeling2021}. The DRP converts raw data into sky-subtracted, flux-calibrated data cubes. In this work, we also use the spectral indices, emission line properties and kinematics derived by the MaNGA Data Analysis Pipeline (DAP) \citep{westfallDataAnalysisPipeline2019, belfioreDataAnalysisPipeline2019}. The DAP analyzes the data from DRP to produce a range of physical properties of the spaxels. Each spaxel is 0.5" in size. The typical point spread function of the MaNGA data is around 2.5" to 3". MaNGA employs dithered observation to produce near-critical sampling of the PSF \citep{lawOBSERVINGSTRATEGYSDSSIV2015}.

In the following section, Gaussian flux, Equivalent Width (EW) of emission lines, and spectral indices provided by the DAP are used.
\section{Methods}
\label{sec: method}
This section consists of four subsections. Section \ref{subsec: spaxel selection} describes the LIERs spaxel selection process. The construction of the corresponding stellar continuum for the selected spaxels is discussed in Sect. \ref{subset: matching}. Section \ref{subsec: driz and stack} talks about the de-redshifting and stacking process. The method used to measure emission line fluxes and estimate uncertainties is described in Sect. \ref{subsec: measurement}.
\subsection{Spaxel selection}
\label{subsec: spaxel selection}
\begin{figure}
        \includegraphics[width=\columnwidth]{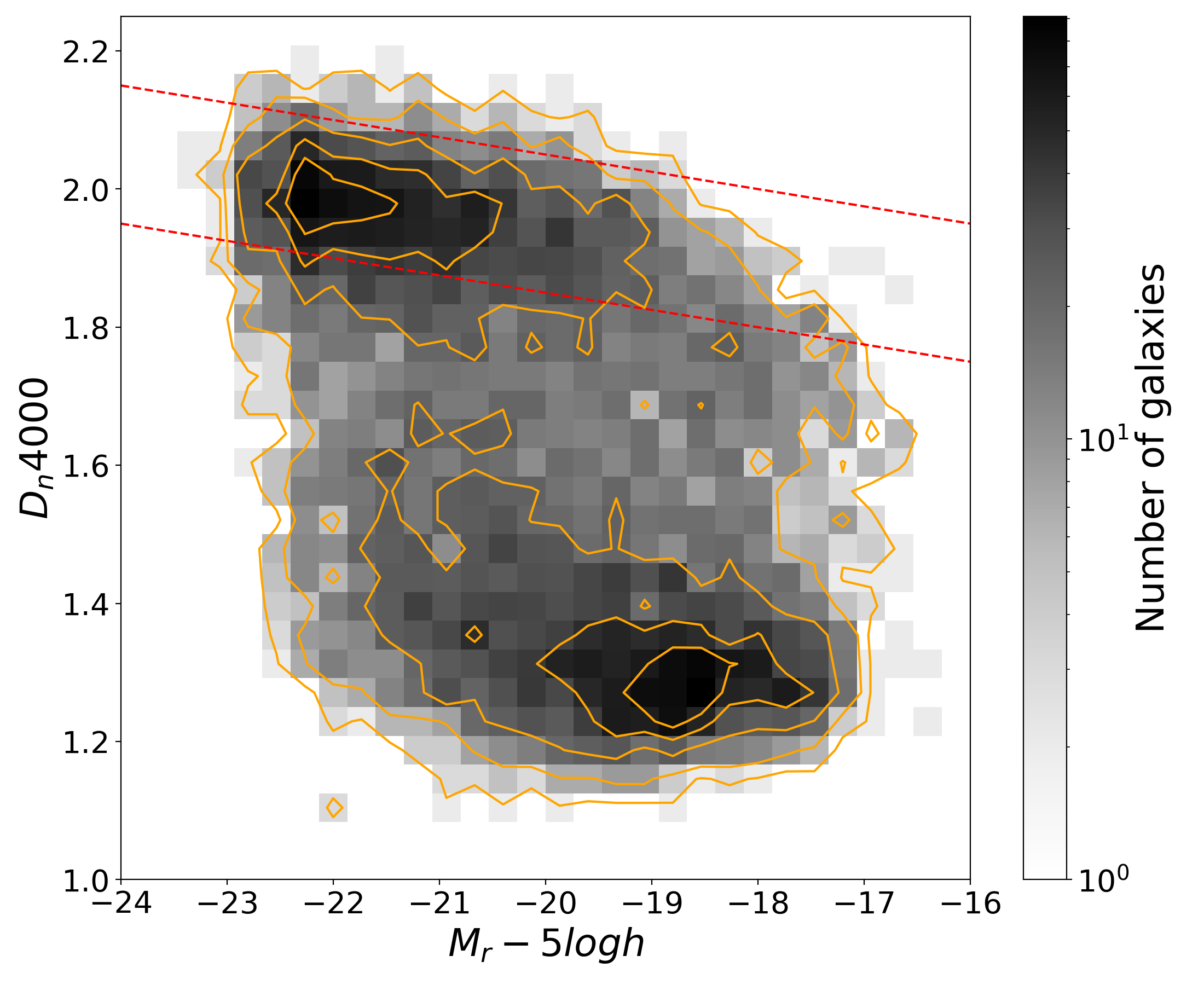}
    \caption{$D_n$4000 vs. absolute magnitude in the $r$ band. The values of $D_n$4000 are from the MaNGA DAP and the absolute magnitudes in the $r$ band are from the NSA catalogue. The gray region shows the density distribution of all galaxies, while the orange contours are drawn at 10\%, 30\%, 68\%, 95\%, and 99\% of all galaxies. Galaxies between the red dashed lines are selected as our quiescent galaxy sample. Galaxies with invalid M$_r$ and D$_n$4000 are not plotted.}
    \label{fig:Galaxy cut}
\end{figure}

To select quiescent galaxies with negligible ongoing star formation, we constructed a plot with $D_n$4000 versus absolute magnitude in the $r$ band, which is shown in Fig. \ref{fig:Galaxy cut}. The absolute magnitudes are from the NASA-Sloan Atlas (NSA) catalogue. $D_n$4000 is a dust-insensitive stellar age indicator, which can help us identify quiescent galaxies without including dusty star-forming galaxies. We adopted the definition of $D_n$4000 from \cite{baloghDifferentialGalaxyEvolution1999}, who define this index as the ratio of average flux density between 4000-4100\AA\ and 3850-3950\AA. In Fig. \ref{fig:Galaxy cut}, the galaxies are divided into two groups, the blue cloud and the red sequence. Located at the bottom right, the blue cloud represents active galaxies hosting young stellar populations. They have lower $D_n$4000 values and on average fainter in the $r$ band. On the contrary, the quiescent galaxies with older stellar populations are at the top left region. 

To select our targeted galaxies, we first drew a line following the ridge-line of the quiescent galaxies distribution. Then, we shifted it upward and downward such that it covered the whole quiescent sequence. The formulae of the boundary lines are given by
\begin{equation}
    D_n4000 > -0.025(M_r - 5 \log h) + 1.35
\end{equation}
and
\begin{equation}
    D_n4000 < -0.025(M_r - 5 \log h) + 1.55
,\end{equation}

We only included spaxels with mean g-band weighted signal-to-noise ratio per pixel greater than 15 in the selected galaxies, yielding a total of 492,275 spaxels. We then applied a zero point correction as described in Sect. 3.3 of \cite{yanShocksPhotoionizationDirect2018}. The rationale is to remove any systematic error in the measurements of small EW by DAP. We defined EW to be
\begin{equation}
    \text{EW} = \int \frac{F_s - F_c}{F_c} d\lambda
    \label{eq:EW}
,\end{equation}
where $F_s$ is the flux of the observed spectrum and $F_c$ is the flux of the continuum. For emission lines, the EW is positive.

\begin{figure*}
    \sidecaption
        \includegraphics[width=12cm]{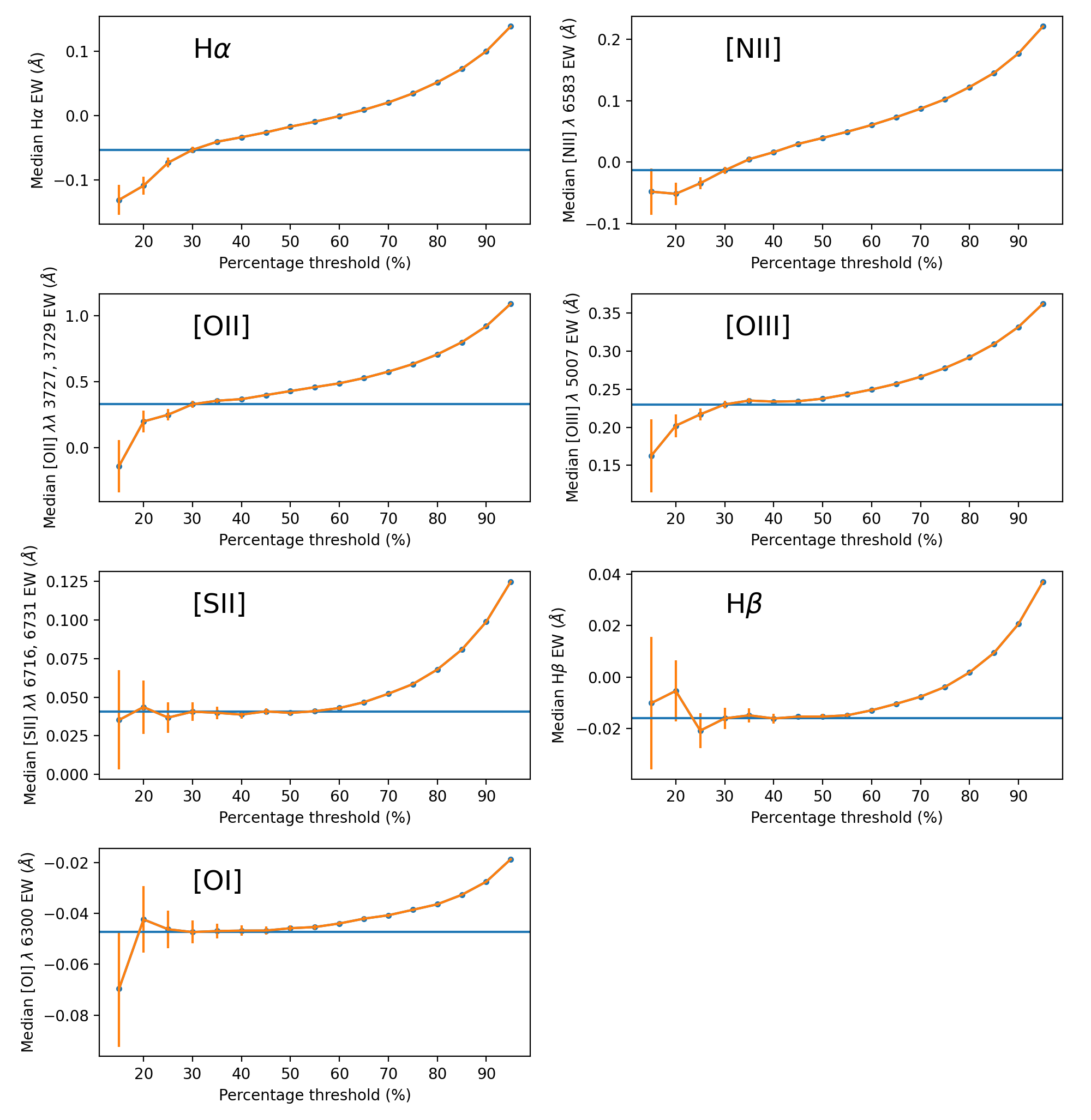}
    \caption{Median EW value of the selected line among spaxels with EW lower than different percentile in other strong lines. The horizontal lines denote the EWs corresponding to 35 percentiles, which are the zero points we chose.}
    \label{fig:Zero point selection}
\end{figure*}

\begin{figure*}
        \includegraphics[width=\linewidth]{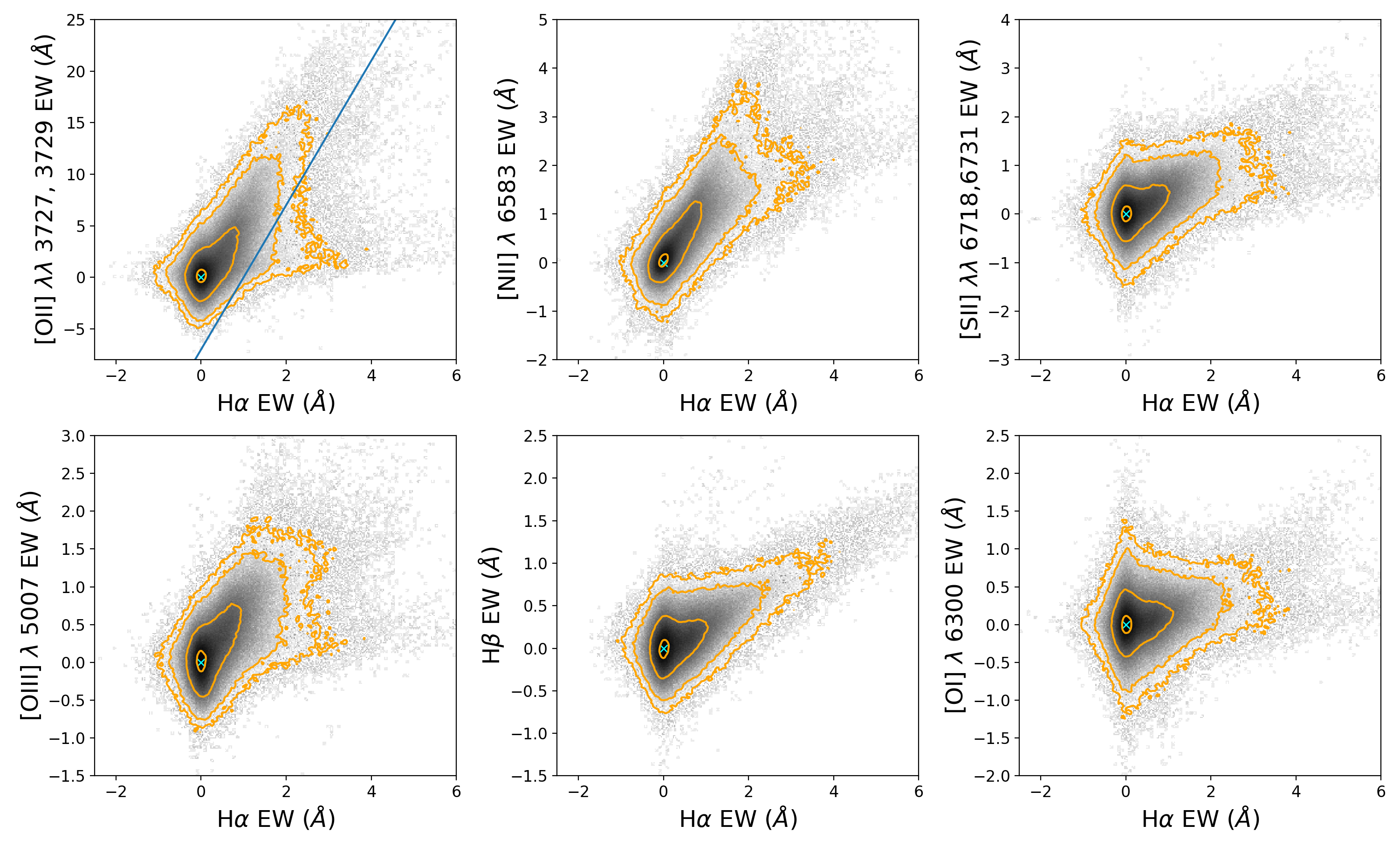}
    \caption{Different strong-line EWs against \ha\ EW after zero point correction. The blue cross denotes the origin. The grayscale represents the density of the spaxels, while the orange contours enclose 10\%, 68\%, 95\%, and 99\% of the spaxels in each panel. The 10\% contour is used to show that most of the data are centered at the zero value. The line in the first subplots represents the cut we applied to exclude contamination from transition objects and dusty star-forming galaxies. All spaxels on the right side of the line are removed from the next step.}
    \label{fig:Zero point corrected}
\end{figure*}

To measure the zero-point of EW, we adopted the assumption that EWs of strong emission lines correlate with each other. This assumption applies because LIERs have fairly uniform line ratios and also very uniform stellar continuum shapes. Here we considered \oii\ $\lambda\lambda$3727, 3729, \nii\ $\lambda$6583, \sii\ $\lambda\lambda$6718, 6731, \oiii\ $\lambda$5007, \hb, and \oi $\lambda$6300. If an emission line EW is consistent with zero, the EW value of each of the other lines of this spaxel should also be among the weakest. To evaluate the real zero-point of a certain emission line, we looked at other lines. For each of the other lines, spaxels are divided into different percentiles according to their EW. Spaxels that lies below a certain percentile in all emission line except the targeted one were selected. The median EW value of the targeted line among the selected spaxels was then plotted against percentiles in Fig. \ref{fig:Zero point selection}. The median EW of the 35th percentile was chosen as the zero-point since it represents the flattest part in most of the plots, showing a stable zero-point without much disruption by noise. Figure \ref{fig:Zero point corrected} shows the distribution of EWs after the correction. One can see from these figures that, after the correction, the most densely populated regions are around (0,0) in all EW versus EW diagrams. Besides correcting the EWs for the zero-point offsets, we also correct the Gaussian emission-line fluxes accordingly.

 As seen in the \oii\ EW versus \ha\ EW panel in Fig. \ref{fig:Zero point corrected}, there is another group of spaxels to the right of the main group. This might be due to contamination from other transition objects and star-forming galaxies which appear red due to dust. According to \cite{yanOriginIiEmission2006}, these objects should be excluded using an \oii-\ha\ EW cut. Hence, we excluded these objects by the  cut 
 \begin{equation}
    \text{EW}(\oii) < 7\text{EW}(\ha) - 7,
 \end{equation}
which is shown in Fig. \ref{fig:Zero point corrected}. This removed 25,526 spaxels with the remaining 466,749.

\begin{figure*}
        \includegraphics[width=\linewidth]{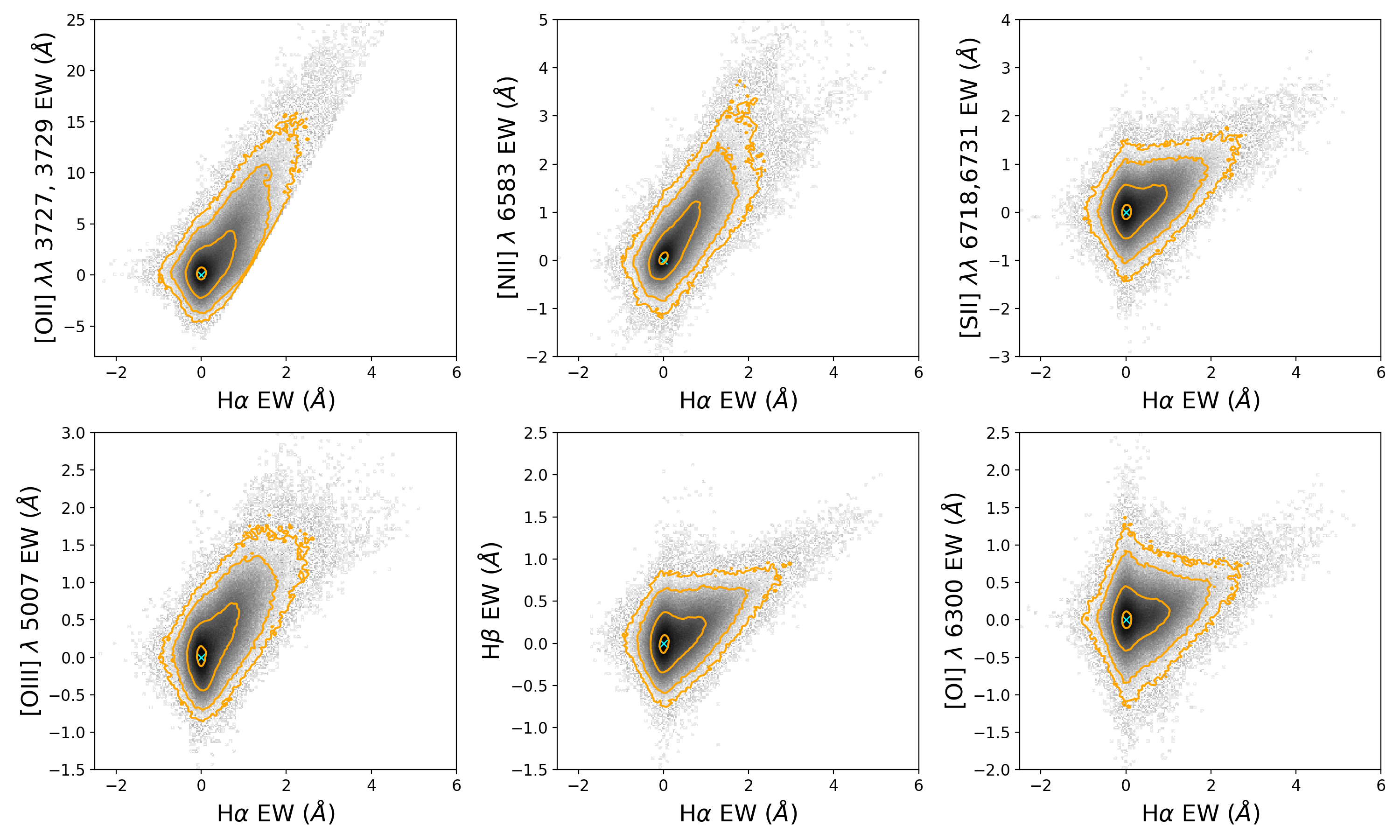}
    \caption{Different strong-line EWs against \ha\ EW after excluding dusty and Seyfert spaxels. The gray regions and contours are generated in the same way as those in Fig. \ref{fig:Zero point corrected}.}
    \label{fig:Zero point exclude}
\end{figure*}

To prevent possible contamination of spaxels coming from Seyfert galaxies, we applied another cut to the sample using the definition of Seyfert galaxies in \cite{heckmanOpticalRadioSurvey1980}. We selected spaxels with a fractional error\footnote{The fractional error a line ratio, A to B, is defined to be $\sqrt{\frac{\sigma_A^2}{I_A^2} + \frac{\sigma_B^2}{I_B^2}}$, where $\sigma$ is the measurement uncertainty and $I$ is the measured flux of each line.} of \oiii/\oii\ flux lower ratio than 0.3 and excluded those with \oiii/\oii\ flux ratio greater than 1. This removed 2676 spaxels and 464,073 spaxels were remaining. The distributions of EWs after removing these dusty and Seyfert spaxels are displayed in Fig. \ref{fig:Zero point exclude}. 

Next, the spaxels are separated into two classes, strong-line samples and zero-line samples. The former contains spaxels with strong emission from the gas and will be regarded as the data to be stacked and measured. Gas emission is very weak in the latter class, making these spaxels suitable candidates for constructing stellar continua. To select strong-line samples, we constructed a parameter named total EW using all EWs from the targeted lines. Here we considered \ha, \oii\ $\lambda\lambda$3727, 3729, \nii\ $\lambda$6583, \sii\ $\lambda\lambda$6718, 6731, and \oiii\ $\lambda$5007 they are the strongest lines in the optical spectrum. The Total EW is defined as follows: 
\begin{equation}
\centering
\begin{split}
    \text{Total EW} =& \text{ EW(\ha)} + 1.15\text{EW(\nii) + 7EW(\oii)} \\
    &+0.55\text{EW(\oiii) + 0.4EW(\sii).}
\end{split}
\end{equation}
The coefficients were chosen to match the slope of the distribution in Fig. \ref{fig:Zero point corrected} so that the Total EW increases along the sequence in each diagram. Spaxels with the highest 20\% of total EW index are often classified as the strong-line sample, and the spaxels are regarded as strong-line spaxels. In the selection, we only included the spaxels with unmasked EW values on all of the above emission lines. There are a total of 90,359 strong-line spaxels.

For a strong-line spaxel to be stacked, we need to know its stellar and gas kinematics, which can differ from each other. We describe how we deal with this in Sect. \ref{subsec: driz and stack}. Therefore, we only included strong-line spaxels with valid line-of-sight stellar velocity and \ha\ velocity from the DAP. We also required the difference between the two velocities to be less than 300 kms$^{-1}$, to remove the most extreme velocity offsets. In total, these selections excluded 2390 strong-line spaxels with 87,969 remaining.  

\begin{figure}
        \includegraphics[width=\columnwidth]{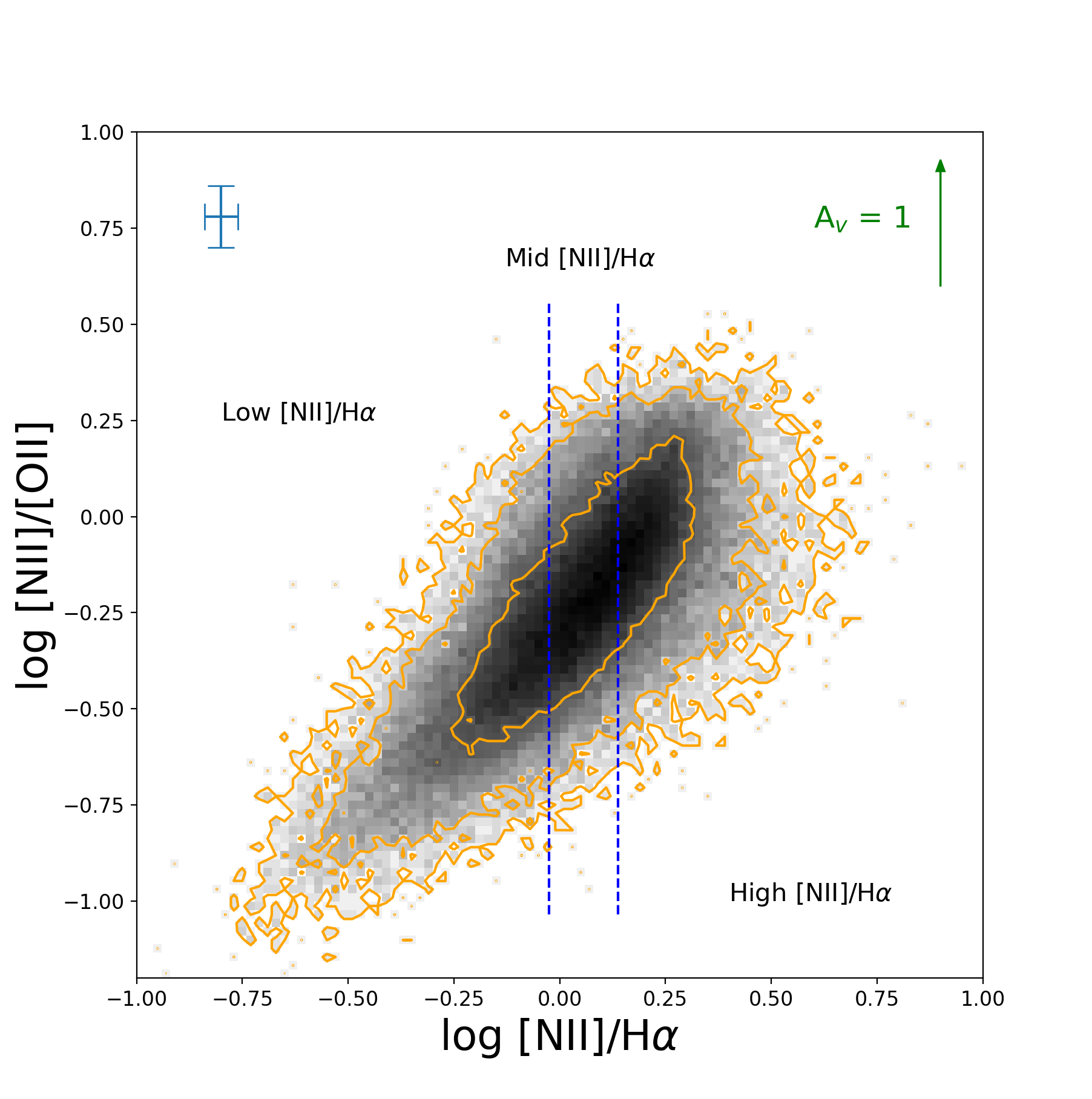}
    \caption{\nii/\oii\ vs. \nii/\ha\ for all spaxels in the strong-line sample. Instead of a narrow sequence, there is noticeable scatter in this relation. The two blue dashed lines separate the spaxels into high, mid, and low \nii/\ha\ bins. The data is meshed into 100$\times$100 bins. The grayscale represents the density of all selected spaxels in each bin, while the orange contours enclose 68\%, 95\%, and 99\% of the spaxels. The blue error bars in the top left corner denote the typical 1$\sigma$ uncertainty of the line ratios. The extinction vector is given at the top right corner.}
    \label{fig:metal}
\end{figure}

Since metals are effective coolants in the ISM, variations in metallicity among different spaxels can cause large discrepancies in measured temperatures. Therefore, we divided the strong-line sample into three bins, namely the high, mid, and low \nii/\ha\ bins. We plotted two metallicity indicators, \nii/\oii\ and \nii/\ha, against each other in Fig. \ref{fig:metal}. Due to the large wavelength difference of \nii\ and \oii\ strong lines, dust extinction can significantly affect the \nii/\oii\ values. Hence, we divided the sample only based on dust insensitive \nii/\ha\ values. We only selected spaxels with a fractional error for \nii/\ha\ flux ratio lower than 0.3. The selected spaxels were then allocated into three equal-number bins, with 28,098 spaxels in the high \nii/\ha\ bin and 28,097 spaxels in each of the remaining bins. The \nii/\ha\ values separating the three bins are -0.0247 and 0.1374, respectively. Although our rationale was to use \nii/\ha\ as a metallicity indicator, it actually may be sensitive to other parameters, such as the ionization parameter in photonionization or shock velocity in the case of shocks. Therefore, the bins are referred to as \nii/\ha\ bins instead of metallicity bins throughout the paper.

Spaxels with negligible gas emission are selected as the zero-line sample which provides the stellar continuum to be subtracted from the spectra with strong gas emission. Here we considered the same emission lines as the strong-line sample selection, which include \ha, \oii\ $\lambda\lambda$3727, 3729, \nii\ $\lambda$6583, \sii\ $\lambda\lambda$6718, 6731, and \oiii\ $\lambda$5007. Following \cite{yanShocksPhotoionizationDirect2018}, we constructed a multi-dimensional ellipsoid around 0 EW. To select points that are consistent with 0, we need to know the typical noise in the measurement. Since we are measuring the EWs of emission lines, all negative EW values should be due to systematics and random errors. We used these negative EW to estimate the typical uncertainty for each emission line by calculating their root-mean-squared (RMS) value from 0. The EW of each line was normalized by this typical uncertainty. We then squared them and summed them up, then took the square root to calculate an uncertainty-normalized distance away from the origin. For zero-line sample selection, we required this distance in the multidimensional space to be less than 2.5. In total, there are 185,604 zero-line spaxels.

\subsection{Matching of strong- and zero-line samples}
\label{subset: matching}
In order to measure the weak auroral lines, we needed an accurate subtraction of the stellar continuum. We selected and stacked spectra from the zero-line sample to construct a stellar continuum matching that of the strong-line stack. In order to get a good matching, we tried to select spaxels from the zero-line sample matching the stellar population of each spaxel in the strong-line sample.
Each strong-line spaxel contains a spectrum composed of a stellar continuum and gas emission lines. We used four proxy parameters, $D_n$4000, stellar velocity dispersion, surface luminosity in the $r$-band, and total spectral broadening to match the stellar continuum shapes. 
\begin{enumerate}
    \item $D_n$4000: The $D_n$4000 values are from the DAP, following the definition mentioned in Sect. \ref{subsec: spaxel selection}. $D_n$4000 is a stellar age and metallicity indicator. Both of them largely affect the shape of the stellar continuum.\\
    \item Stellar velocity dispersion ($\sigma_\ast$): DAP provides the measurement of the line-of-sight stellar velocity dispersion. Because $\sigma_\ast$ correlates with the stellar metallicity [Fe/H], and especially [Mg/Fe] \citep{gravesDissectingRedSequence2009, bernardiEarlyTypeGalaxiesSloan2003}, matching it can help match the stellar continuum.\\
    \item Surface luminosity in the $r$-band: The $r$-band flux per pixel is from the DRP. We corrected for the galactic reddening effect using the E(B-V) value given by the DAP.  
    As all spaxels in our sample have similar colors, matching the surface luminosity in $r$-band can help match the stellar mass surface density between spaxels, which can facilitate the matching of properties affecting the stellar continuum.\\
    \item Total spectral broadening: Total spectral broadening is the quadratic sum of stellar velocity dispersion $\sigma_\ast$ and the instrumental line spread function (LSF). The instrumental LSF values are the post-pixelized LSF given by the DRP. We first convert the post-pixelized LSF value into velocity space, then take the median across the whole spectrum before combining with $\sigma_\ast$. After that, we convert back the median value into wavelength space and quadratically summed it with $\sigma_\ast$. Matching the total spectral broadening matches the absorption line width between the strong and zero-line spaxels.
\end{enumerate}
We excluded strong and zero-line spaxels with any of the above parameters masked by the DRP or DAP. There are 183,736 zero-line spaxels. The high, mid, and low \nii/\ha\ bins of the strong-line sample have 27,715, 27,883, and 27,924 spaxels, respectively. Next, we normalized these four parameters using their 5th and 95th percentiles as 0 and 1, respectively. We used a K-Dimensional tree for matching the spaxels. A four-dimensional space was created using the four normalized parameters. For each strong-line spaxel, we found the 2 non-occupied zero-line spaxels that are closest to it in the 4-D space. In each iteration, each strong-line spaxel is matched with the two closest zero-line spaxels. It is possible that a zero-line spaxel will be matched with more than one strong-line spaxel. In such a case, only the matching between the zero-line spaxel and its closest strong-line spaxel will be kept. Strong-line spaxels with fewer than two matches will be kept for the next iteration. The maximum iteration is 25, and the remaining strong-line spaxels will be abandoned. After all the iterations, the matching distance is required to be less than 0.05 in the four-dimensional space, or the trio will be excluded. There are 25,139, 26,435 and 25,229 spaxel trios in the high, mid, and low \nii/\ha\ bins respectively.

\subsection{Drizzling and stacking}
\label{subsec: driz and stack}
The position of a certain emission line in the spectrum depends on the redshift of the source spaxel. Within the region covered by the spaxel, ionized gas velocity can be different from the stellar kinematics. Therefore, the stellar absorption lines and the gas emission lines are shifted by different extents. To detect emission lines in the stacked spectrum of each \nii/\ha\ bin, the spectra of the spaxels must be shifted according to their gas kinematics. In order to get purely gas emission spectra, the stellar continuum of the strong-line spaxel has to be subtracted. Using the zero-line stacked spectrum as a stellar continuum template, the zero-line spectrum has to be shifted to the same position as the strong-line spaxels' stellar continua. 

\begin{figure*}
\centering
    \includegraphics[width=\linewidth]{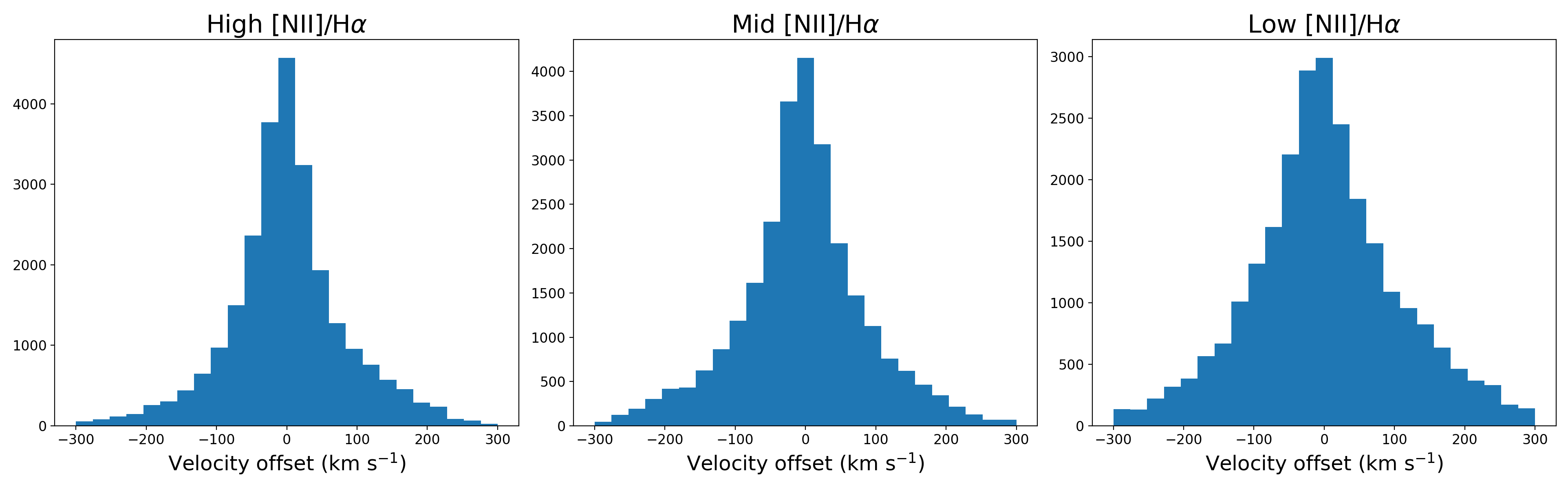}
    \caption{Distribution of velocity offsets of all strong-line spaxels in each \nii/\ha\ bin. The velocity offset is defined as the difference between the stellar velocity and the gas velocity. There are a total of 25 velocity-offset bins covering the range from -300 km s$^{-1}$ to 300 km s$^{-1}$.}
    \label{fig:voffset}
\end{figure*}

We defined a velocity offset as the difference between the stellar and ionized gas velocities. The redshift value, stellar velocity, and gas velocity are all from the DAP. The \ha\ velocities are used as the gas velocities of the spaxels. We divided all the strong-line spaxels into 25 bins of velocity offset ranging from -300 km s$^{-1}$ to 300 km s$^{-1}$. The distributions of strong-line spaxels in the velocity offset bins are given in Fig. \ref{fig:voffset}. To align with the strong-line spaxel, the matched zero-line spectra are first shifted by their redshifts and stellar kinematics. Then they are shifted by the center value of the velocity offset of the velocity-offset bin in which its associated strong-line spaxel is located. In practice, the shifts were done together to minimize errors introduced by the shifts. The overall redshift value is given by Eq.\ \ref{eq:redshift},
\begin{equation}
    \centering
    z_p = (1+(\frac{v_{\ast}-v_{\text{off}}}{c}))(1+z)-1
    \label{eq:redshift}
,\end{equation}
where $z_p$ is the overall redshift, $v_{\ast}$ is the stellar velocity, $v_{\text{off}}$ is the center value of the velocity offset bin and $z$ is the redshift of the galaxy. The term $v_{\ast}-v_{\text{off}}$ means that the spaxels are aligned by the gas velocity, which corresponds to the positions of the emission lines in the spectrum.

All spectra are first corrected for galactic extinction using the extinction curve of \cite{fitzpatrickCorrectingEffectsInterstellar1999} with $R_V$ = 3.1 before the drizzling. MaNGA galaxies are nearby with redshift values smaller than 0.15. Since the covered wavelength is up to 10354 \AA, the maximum wavelength after deredshifting is set to be 8500 \AA\  which allows us to measure all the targeted emission lines. We stored the shifted spectra in a common wavelength grid, evenly spaced logarithmically with the same spacing as the MaNGA logcube. The process of assigning flux among the output pixel is referred to as drizzling.

In the following context, a pixel refers to the region between two wavelengths in the wavelength array of the MaNGA logcube. The re-distribution of flux among the output pixels should conserve flux locally. In the drizzling process, the original wavelength array should shrink by a factor of $(1+z)$, and the flux in each pixel is multiplied by this amount to conserve the total flux. The shifted pixel could fall across the boundary of two pixels in the output array. We allocated the flux of each shifted pixel according to its overlapping fractions with the final pixel grid. Since both the input and output array are logarithmic, the proportion depends only on the $z_p$ value of the spaxel and it remains the same across the whole wavelength array.

\begin{figure*}
    \sidecaption
        \includegraphics[width=12cm]{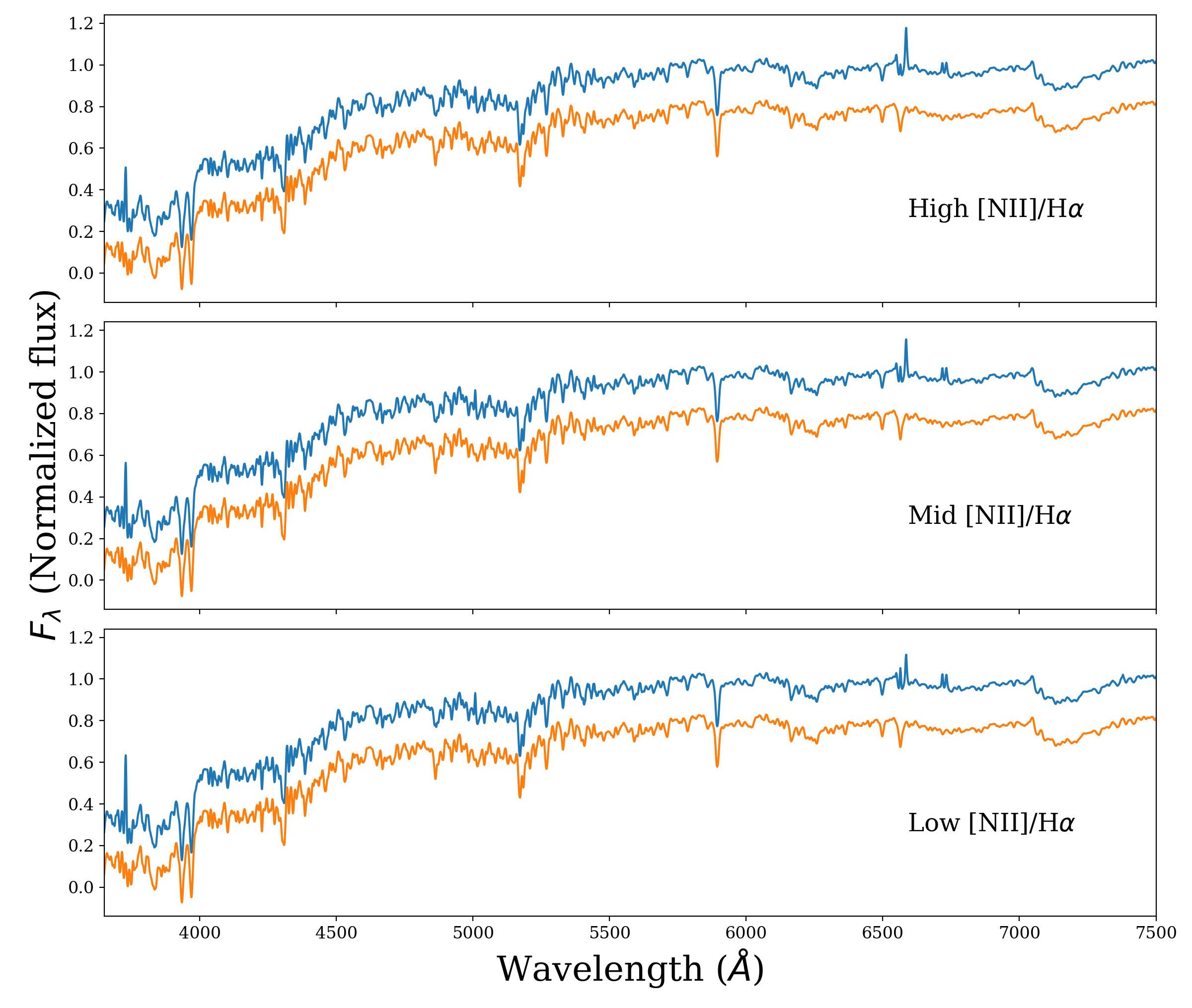}
    \caption{Stacked strong-line spectrum (blue) and zero-line spectrum (orange) of the three \nii/\ha\ bins. The zero-line spectrum has been corrected for overall shape using B-spline and is vertically shifted downward for visibility.}
    \label{fig:strong_control}
\end{figure*}

\begin{figure*}
    \sidecaption
        \includegraphics[width=12cm]{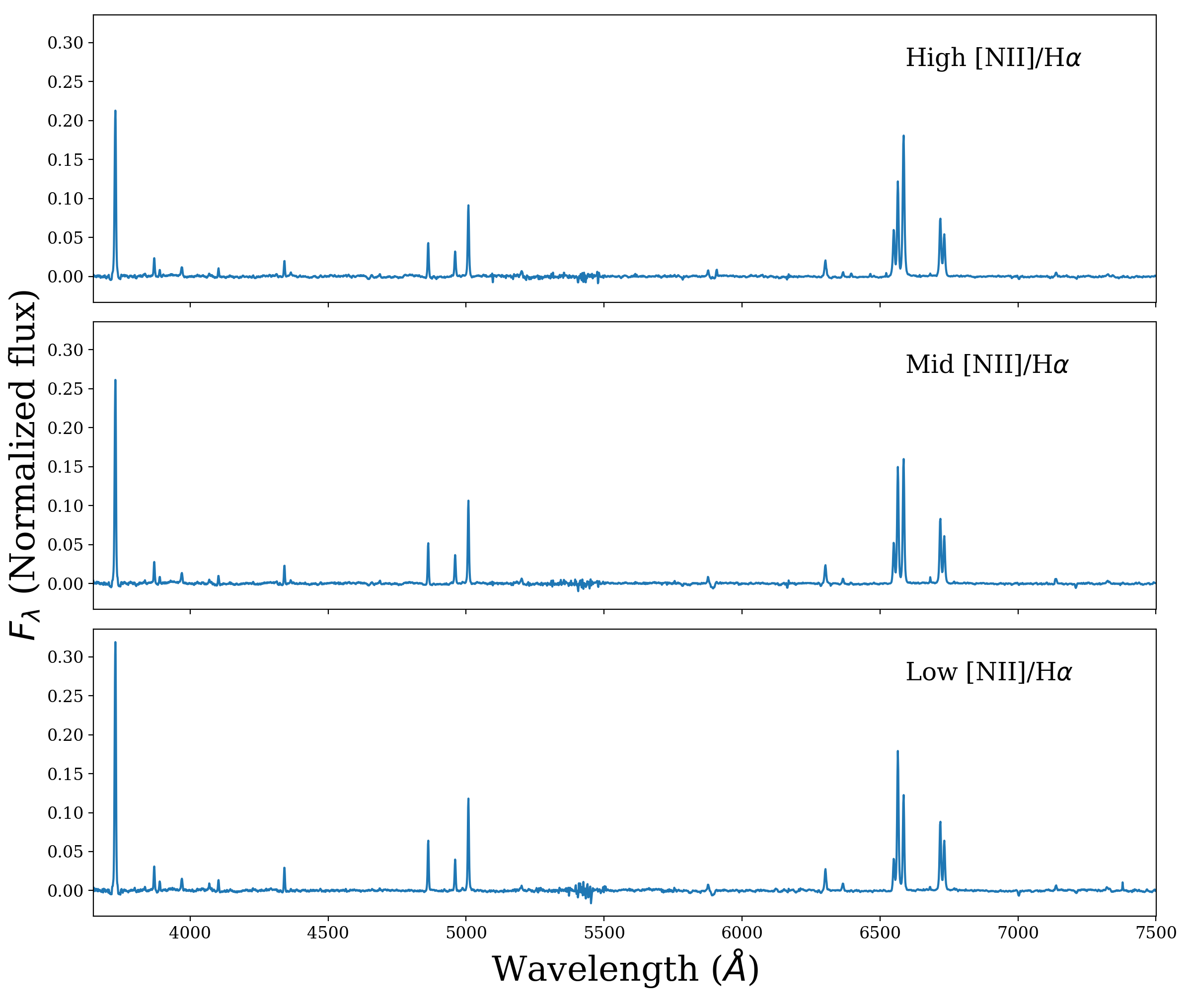}
    \caption{Continuum-subtracted residual spectrum of the three \nii/\ha\ bins. All the lines in the spectrum should originate from gas emissions.}
    \label{fig:residual}
\end{figure*}

All the spectra after de-redshifting are normalized by their median flux values of the region between 6000 - 6100\AA. Within each \nii/\ha\ bin, strong-line spaxels and zero-line spaxels in each velocity offset bin are stacked together respectively. Similar to \cite{yanShocksPhotoionizationDirect2018}, there are small discrepancies between the overall shapes of the stacked strong-line and zero-line spectra. We followed the same procedure to solve the issue. We first divided the strong-line spectrum by its corresponding zero-line spectrum and applied a B-spline fitting on the resulting curve with a spacing of 400 pixels between the breakpoints. Emission lines are masked in the fitting as the goal is just to get the overall shape difference. The stacked zero-line spectrum is then multiplied by the B-spline curve to match the shape of the stacked strong-line spectrum. The gas emission spectra are obtained by subtracting the shape-corrected zero-line spectrum from the strong-line spectrum. Finally, the gas emission spectra of all velocity offset bins are stacked together, weighted by the number of spectra stacked in that velocity offset bin. Finally, the gas emission spectrum of each of the three \nii/\ha\ bins is obtained. Figure \ref{fig:strong_control} shows the strong-line and zero-line spectra of the three bins. The gas emission spectra are displayed in Fig. \ref{fig:residual}.

\subsection{Emission line measurement}
\label{subsec: measurement}
We followed the same method in \cite{yanShocksPhotoionizationDirect2018} and \cite{yanOriginIiEmission2006} for flux measurement. Two sidebands and a central region are selected for each line. A linear fit of the sidebands is used as an estimation for the local continuum shape and is subtracted before integrating the flux inside the central region. We divided the emission lines into two groups, strong lines and weak lines. The definition of strong and weak lines with the corresponding regions defined can be found in Table \ref{table: strong def} and \ref{table: weak def}. Assuming all the lines have a symmetric profile, the central region is defined to be centered at the vacuum wavelength of the line in most of the singlet cases. A special case is \he, which is treated as a doublet with the coinciding \neiii\ $\lambda$3968. For the doublets, the region extends to the left of the bluer line with the same amount to the right of the redder line. For \nii\ $\lambda\lambda$6548, 6583 and \ha, summed flux is not used and the regions are only defined to estimate the noise and fit the Voigt profile as described in the next paragraph. By visual inspection, the central region can cover the whole emission line profile in all of the lines.

\begin{table*}
\caption{Definition of the central region and sidebands for strong lines}
\label{table: strong def}
\centering
    \begin{tabularx}{\linewidth}{@{}l *4{>{\centering\arraybackslash}X}@{}}
    \hline
    Line              & Central region (\AA)   & Left sideband (\AA)   & Right sideband (\AA)   \\
    [2pt]
    \hline
    \oii\ $\lambda\lambda$3727, 3729  & 3717.092 - 3739.875 & 3667.092 - 3717.092 & 3739.875 - 3789.875 \\
    [3pt]\neiii\ $\lambda$3869      & 3859.85 - 3879.85   & 3839.85 - 3859.85   & 3898.85 - 3912.85\\
    [3pt]\hb            & 4852.72 - 4872.72   & 4802.72 - 4852.72   & 4872.22 - 4922.22    \\
    [3pt]\oiii\ $\lambda$5007       & 4998.24 - 5018.24   & 4968.24 - 4998.24   & 5018.24 - 5048.24    \\
    [3pt]\oi\ $\lambda$6300         & 6287.046 - 6317.046 & 6252.046 - 6287.046 & 6317.046 - 6352.046  \\
    [3pt]\ha           & 6554.6 - 6574.6   & 6484.85 - 6534.85   & 6600.27 - 6650.27    \\
    [3pt]\nii\ $\lambda$6583        & 6570.27 - 6600.27   & 6484.85 - 6534.85   & 6600.27 - 6650.27    \\
    [3pt]\sii\ $\lambda\lambda$6716, 6731 & 6703.29 - 6748.67   & 6653.29 - 6703.29   & 6748.67 - 6798.67    \\
    \hline
    \end{tabularx}
\end{table*}

\begin{table*}
\caption{Definition of the central region and sidebands for weak lines}
\label{table: weak def}
    \begin{tabularx}{\linewidth}{@{}l *4{>{\centering\arraybackslash}X}@{}}
    \hline
     Line             & Central region (\AA)   & Left sideband (\AA)   & Right sideband (\AA)   \\
    [2pt]
    \hline
     \hz           & 3882.166 - 3898.166 & 3840 - 3860         & 3898.166 - 3918.166  \\
     [3pt]\he        & 3960.59 - 3979.198  & 3940.59 - 3960.59   & 3979.198 - 3999.198  \\
     [3pt]\sii\ $\lambda\lambda$4068, 4076 & 4061.75 - 4085.5    & 4052.75 - 4061.75   & 4085.5 - 4094.5      \\
     [3pt]\hd          & 4094.892 - 4110.892 & 4086.892 - 4094.892 & 4110.892 - 4118.892  \\
     [3pt]\hg          & 4333.692 - 4349.692 & 4323.692 - 4333.692 & 4349.692 - 4359.692  \\
     [3pt]\oiii\ $\lambda$4363      & 4356.436 - 4372.436 & 4303.436 - 4333.436 & 4372.436 - 4402.436  \\
     [3pt]\nii\ $\lambda$5755       & 5748.119 - 5764.119 & 5668.119 - 5748.119 & 5764.119 - 5844.119  \\
     [3pt]\hei\ $\lambda$5876       & 5869.24 - 5885.24   & 5818 - 5868         & 5915 - 5965          \\
     [3pt]\oii\ $\lambda\lambda$7320, 7330 & 7313.94 - 7340.21   & 7283.94 - 7313.94   & 7340.21 - 7370.21    \\
     [3pt]\heii\ $\lambda$4687      & 4677 - 4697         & 4617 - 4677         & 4697 - 4757          \\
    \hline
    \end{tabularx}
\end{table*}

Since \nii\ $\lambda\lambda$6548, 6583 are overlapping with \ha, we estimate the fluxes of the three lines with pseudo-voigt fits. The pseudo-voigt profile is a linear combination of a Gaussian and a Lorentzian profile which has one more degree of freedom in the fitting process than the Gaussian and Lorentzian profiles. We tried to fit all three profiles and the pseudo-voigt has the best estimation for the shape of the lines as it has one more degree of freedom than the other two profiles. Before the fitting process, the spectrum is first subtracted with the linear fit estimated from the sidebands. Singlets are fitted only by providing the vacuum wavelength as the initial guess. For the doublets, the stronger line position is provided as the initial guess, and the relative position of the other line is fixed. For \nii\ and \ha, the three line profiles are fitted simultaneously as a combination of a singlet and a doublet. The relative strength of the two \nii\ lines has a fixed ratio as required by quantum mechanics, where $I_{6583}/I_{6548} = 3.071$ \citep{storeyTheoreticalValuesOIII2000} and $I$ is the flux intensity of the line. In our following discussion, only the flux from \nii\ $\lambda$6583 is considered. The flux of \ha\ and \nii\ $\lambda$6583 used are both estimated by integrating the pseudo-Voigt profile, which is referred to as the Voigt flux. The measured Voigt fluxes of other strong lines with comparison to the summed fluxes are provided in Table \ref{table: voigt}. All the fluxes are normalized relative to the summed flux of \hb. It can be seen that the pseudo-Voigt profiles can capture the fluxes of the lines well to better than 3\% assuming the summed flux is the true line flux.  

\begin{table*}
    \caption{Summed flux and Voigt flux comparison for the strong lines other than \nii\ and \ha. All the fluxes are normalized by the summed flux of \hb.}
    \begin{subtable}{
        \begin{tabularx}{\linewidth}{@{}l *7{>{\centering\arraybackslash}X}@{}}
        \hline
         High \nii/\ha   &   \oii\ $\lambda\lambda$3727, 3729 &   \neiii\ $\lambda$3869 &   \hb &   \oiii\ $\lambda$5007 &   \oi\ $\lambda$6300 &   \sii\ $\lambda\lambda$6716, 6731 \\
        [2pt]
        \hline
        Summed                  &            651.048 &         52.593 &  100     &       250.06  &      83.827 &            469.794 \\
        [3pt]Voigt                   &            644.991 &         53.337 &  100.309 &       248.688 &      82.732 &            474.667 \\
        [3pt]Flux difference (\%)     &             -0.93  &          1.415 &    0.309 &        -0.549 &      -1.306 &              1.037 \\
        \hline
        \end{tabularx}}
    \end{subtable}
    \begin{subtable}{
        \begin{tabularx}{\linewidth}{@{}l *7{>{\centering\arraybackslash}X}@{}}
         Mid \nii/\ha   &   \oii\ $\lambda\lambda$3727, 3729 &   \neiii\ $\lambda$3869 &   \hb &   \oiii\ $\lambda$5007 &   \oi\ $\lambda$6300 &   \sii\ $\lambda\lambda$6716, 6731 \\
        [2pt]
        \hline
        Summed                 &            661.913 &         50.783 &   100    &       237.851 &      74.657 &            428.071 \\
        [3pt]Voigt                  &            656.958 &         51.327 &   100.220 &       236.86  &      73.475 &            430.737 \\
        [3pt]Flux difference (\%)    &             -0.749 &          1.071 &     0.220 &        -0.417 &      -1.583 &              0.623 \\
        \hline
        \end{tabularx}}
    \end{subtable}
    \begin{subtable}{
        \begin{tabularx}{\linewidth}{@{}l *7{>{\centering\arraybackslash}X}@{}}
         Low \nii/\ha   &   \oii\ $\lambda\lambda$3727, 3729 &   \neiii\ $\lambda$3869 &   \hb &   \oiii\ $\lambda$5007 &   \oi $\lambda$6300 &   \sii\ $\lambda\lambda$6716, 6731 \\
        [2pt]
        \hline
        Summed                 &            605.476 &         42.891 &  100     &       197.803 &      61.307 &            332.048 \\
        [3pt]Voigt                  &            601.369 &         44.013 &  100.366 &       197.062 &      60.882 &            333.398 \\
        [3pt]Flux difference (\%)    &             -0.678 &          2.616 &    0.366 &        -0.375 &      -0.693 &              0.407 \\
        \hline
        \end{tabularx}}
    \end{subtable}
    \label{table: voigt}
\end{table*}

The uncertainties of the measurements are estimated by sliding boxes in the sidebands. After the correction of the linear fitline, the spectrum is locally flat with tiny fluctuations. We assumed the fluctuations in the central region are the same as in the sidebands. We slid a box, which has the same size as the central region, across the sidebands and measured the summed flux inside the box. The position of the box is shifted by one pixel each time and the RMS flux value of all the boxes is taken as the uncertainty of the measurement. For some of the lines, narrower sidebands are chosen to capture the local continuum shape. For lines with narrow sidebands in which fewer than 20 boxes can be fitted, the uncertainty is calculated by first computing the RMS value of all pixels in the sidebands, then multiply it with $\sqrt{N}$, where $N$ is the number of pixels in the central region. The uncertainties and the fluxes of all the lines are given in Table \ref{table: flux_uncer}. Among them, the fluxes of \nii\ and \ha\ are given by the pseudo-Voigt fit. 

\begin{table*}
    \caption{All the fluxes with their uncertainties. All the fluxes are normalized by \hb\ fluxes.}
    \begin{tabularx}{\linewidth}{@{}l *4{>{\centering\arraybackslash}X}@{}}
    \hline
     Lines            & High \nii/\ha   & Mid \nii/\ha   & Low \nii/\ha   \\
     [2pt]
    \hline
      \ha          & 377.4$\pm$5.7     & 411.1$\pm$1.6    & 377.4$\pm$1.4    \\
    [3pt]\hb           & 100.0$\pm$4.4     & 100.0$\pm$3.0    & 100.0$\pm$2.0    \\
    [3pt]\hg          & 40.5$\pm$0.8      & 39.6$\pm$1.2     & 41.1$\pm$0.8     \\
    [3pt]\hd          & 21.4$\pm$0.8      & 18.5$\pm$0.8     & 18.0$\pm$0.7     \\
    [3pt]\he        & 36.0$\pm$1.2      & 32.2$\pm$0.4     & 25.6$\pm$0.6     \\
    [3pt]\hz           & 10.0$\pm$1.2      & 8.0$\pm$1.2      & 12.3$\pm$1.3     \\
    [3pt]\oii\ $\lambda\lambda$3727, 3729 & 651.0$\pm$8.9     & 661.9$\pm$8.6    & 605.5$\pm$6.5    \\
    [3pt]\oiii\ $\lambda$5007      & 250.1$\pm$2.4     & 237.9$\pm$2.3    & 197.8$\pm$2.6    \\
    [3pt]\oi\ $\lambda$6300        & 83.8$\pm$2.3      & 74.7$\pm$2.6     & 61.3$\pm$2.8     \\
    [3pt]\nii\ $\lambda$6583       & 715.9$\pm$7.5     & 496.1$\pm$2.0    & 277.6$\pm$1.3    \\
    [3pt]\sii\ $\lambda\lambda$6716, 6731 & 469.8$\pm$1.6     & 428.1$\pm$2.5    & 332.0$\pm$1.1    \\
    [3pt]\neiii\ $\lambda$3869     & 52.6$\pm$1.3      & 50.8$\pm$1.2     & 42.9$\pm$1.0     \\
    [3pt]\hei\ $\lambda$5876       & 18.3$\pm$2.4      & 18.2$\pm$1.2     & 17.0$\pm$2.1     \\
    [3pt]\heii\ $\lambda$4687      & 8.7$\pm$5.0       & 8.4$\pm$2.9      & 2.7$\pm$1.8      \\
    [3pt]\sii\ $\lambda\lambda$4068, 4076 & 22.1$\pm$1.2      & 21.7$\pm$0.9     & 17.7$\pm$1.2     \\
    [3pt]\oiii\ $\lambda$4363      & 15.2$\pm$2.5      & 9.2$\pm$3.1      & 0.5$\pm$2.0      \\
    [3pt]\nii\ $\lambda$5755       & 3.4$\pm$4.0       & 5.1$\pm$3.6      & 2.5$\pm$3.0      \\
    [3pt]\oii\ $\lambda\lambda$7320, 7330 & 18.0$\pm$1.7      & 17.7$\pm$0.7     & 15.9$\pm$0.8     \\
    \hline
    \end{tabularx}
    \label{table: flux_uncer}
\end{table*}

In the stacking process, the spectra are normalized using the region of 6000-6100 \AA. Therefore, the flux ratio between two lines A and B in the stacked spectrum becomes a weighted average of the per-spaxel line ratios with the weight proportional to the ratio of the line flux in the denominator to the median flux of 6000-6100\AA. This effect can be illustrated using the following equation:
\begin{equation}
    \frac{F_A}{F_B} = \frac{\Sigma_i \frac{F_{iA}}{C_i}}{\Sigma_i \frac{F_{iB}}{C_i}} = \frac{\Sigma_i \frac{F_{iA}}{F_{iB}}\frac{F_{iB}}{C_i}}{\Sigma_i\frac{F_{iB}}{C_i}}
.\end{equation}
Here $F_A$ and $F_B$ are the flux from lines A and B, and $C_i$ is the continuum value used for normalization. We refer to Sect. 4.3 of \cite{yanShocksPhotoionizationDirect2018} for further details.

\begin{figure*}
    \sidecaption
        \includegraphics[width=12cm]{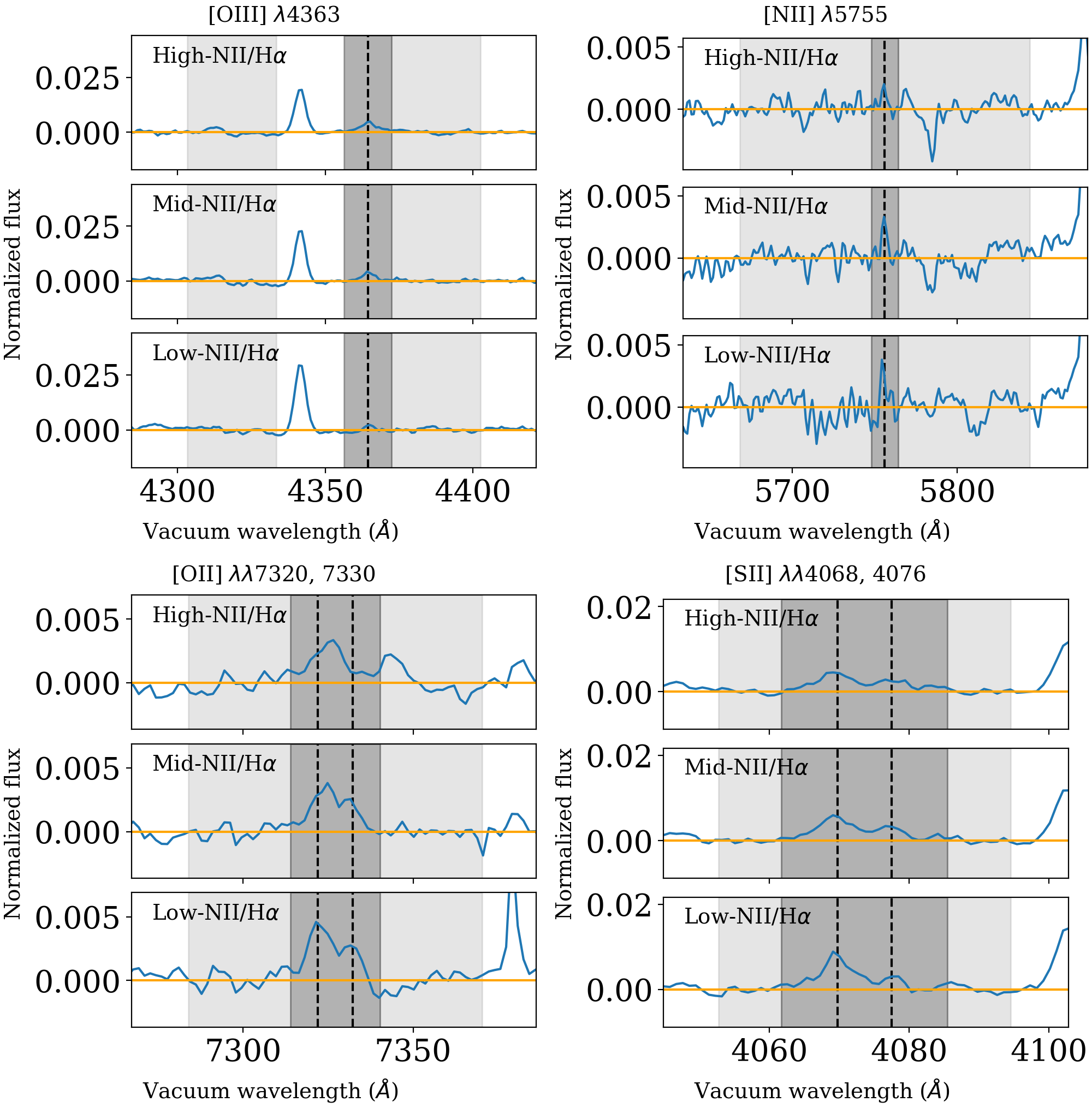}
    \caption{Spectra around the four targeted auroral lines corrected by the fit line derived from the sidebands. The dark gray region is the central region, the light gray regions are the sidebands, and the orange line denotes the zero value.  \oiii\ $\lambda$4363 is on the top left, \nii\ $\lambda$5755 is on the top right, \oii\ $\lambda\lambda$7320, 7330 is at the bottom left, and \sii\ $\lambda\lambda$4068, 4076 is at the bottom right.}
    \label{fig:auroral}
\end{figure*}

Figure \ref{fig:auroral} shows the continuum and emission line of the four targeted auroral lines. Since the auroral lines are weak, in this study a 2$\sigma$ detection threshold is used. Shown in the bottom left panel, \oii\ $\lambda\lambda$7320, 7330 can be well detected in all the \nii/\ha\ bins. Moreover, on the right end of the plotting range, a nickel line \niii\ $\lambda$7378 is observed, especially in the low \nii/\ha\ spectrum. A little bump is also observed in two other bins but their levels are close to the noise level in that region. Shown in the bottom right panel, \sii\ $\lambda\lambda$4068, 4076 is also 2$\sigma$ detected in all the bins. The sidebands of this doublet are chosen to be narrower to capture the local shape of the continuum. The widths of the sidebands of the two lines mentioned are limited by \niii\ $\lambda$7378 and \hd\ respectively. In the top left panel, the flux of \oiii\ $\lambda$4363 increases with increasing metallicity. Although in the low \nii/\ha\ plot, the continuum seems to be lower than 0 which leads to underestimation of flux, a narrower band was used and the result remains the same. As a result, \oiii\ $\lambda$4363 is detected in the high and mid \nii/\ha\ bins but not the remaining bin. \nii\ $\lambda$5755 is not detected in all of the bins despite showing a little bump. The continuum appears to be noisier in that region because the wavelength of the line, when redshifted, lies in the transition region of the blue and red spectrographs in MaNGA. The throughput curves can be found in \cite{yanSDSSIVMaNGASPECTROPHOTOMETRIC2015}.

\section{Data analysis}
\label{sec: data analysis}
In this section, we discuss several physical properties deduced from the measured line ratios. In Sect. \ref{subsec: balmer} we compute the dust extinction coefficient using the Balmer decrement. In Sect. \ref{subsec: density} we use \sii\ doublets to estimate the electron density. Last but not least, temperatures of different ions are computed in Sect. \ref{subsec: temperature}.
\subsection{Balmer decrement and dust extinction}
\label{subsec: balmer}
\begin{figure}
        \includegraphics[width=\columnwidth]{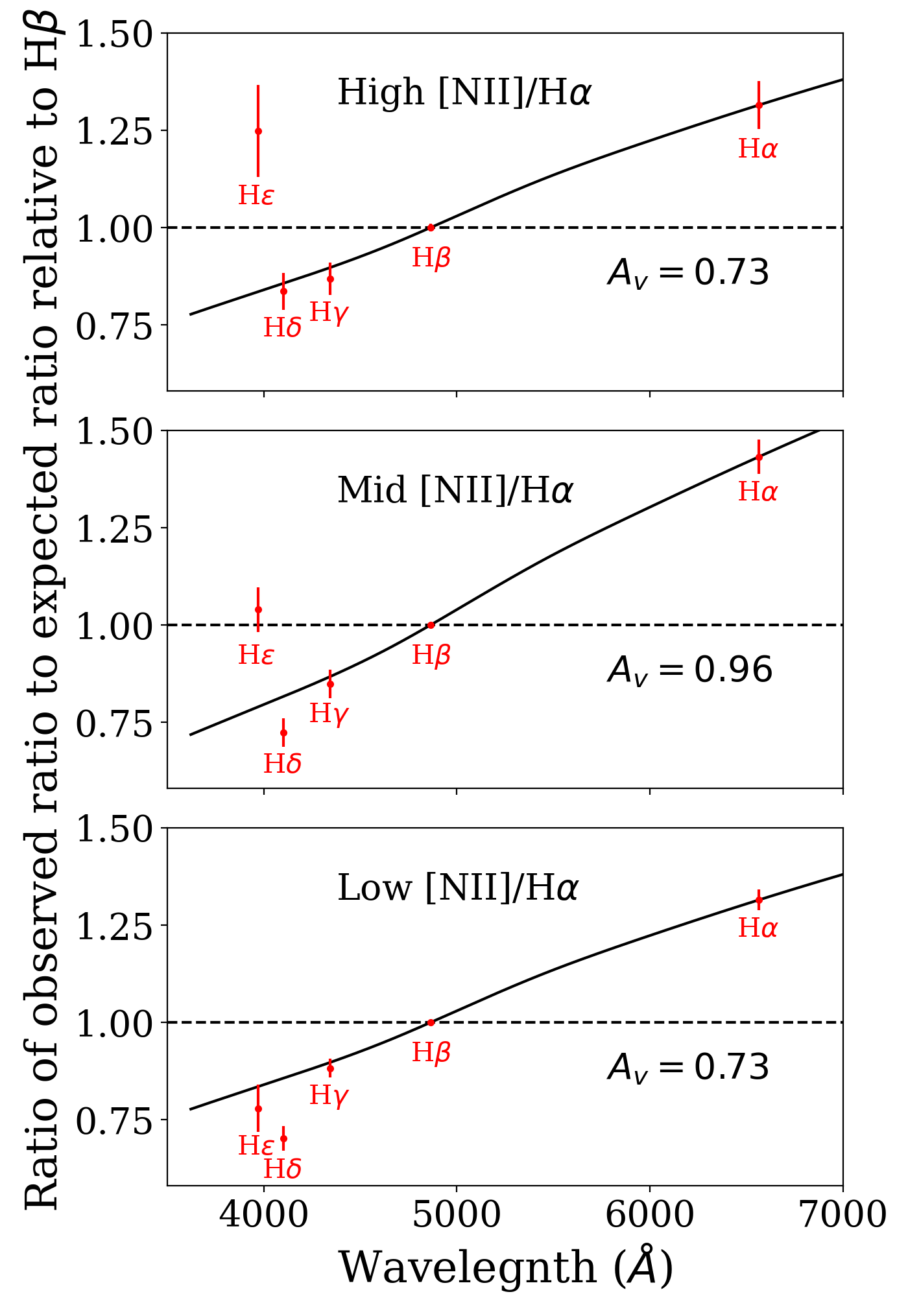}
    \caption{Dust extinction value predicted by the Balmer decrement under case B approximation and low-density limit at 10000 K. While \ha, \hb, \hg, and \hd\ are more consistent, \he\ is consistent with the prediction only in the low \nii/\ha\ bin. \hz\ values are not shown due to strong contamination and stellar continuum systematics.}
    \label{fig:balmer}
\end{figure}

Under case B approximation, the intrinsic ratios of Balmer lines are fixed at certain values at a given electron temperature and density \citep{osterbrockAstrophysicsGaseousNebulae2006}. The observed ratios between the Balmer lines can be used to estimate the degree of dust extinction by comparing it with the theoretical value. We used the extinction curve from \cite{fitzpatrickCorrectingEffectsInterstellar1999} with $R_V$ = 3.1. In Fig. \ref{fig:balmer}, we plotted the ratios of the observed Balmer ratios to the theoretical ratios as a function of wavelength for the first five Balmer lines. Using \ha/\hb\ ratio under case B approximation and at low-density limit at 10,000K, the total extinction (A$_v$) were measured to be 0.73, 0.96, 0.73 magnitudes for high, mid, and low \nii/\ha\ bins, respectively. The black curves in the figure show the predicted values for other wavelengths given these measured extinctions. The red points with error bars are the measured Balmer line fluxes relative to the Case-B values. \he\ overlaps with \neiii\ $\lambda$3967. Since the flux ratio of \neiii\ $\lambda$3967 and \neiii\ $\lambda$3869 is locked by quantum mechanics, we subtracted 31\% of the flux of \neiii\ $\lambda$3869 from the measured \he\ flux. In the same way, part of \hz\ overlaps with \hei\ $\lambda$3889 and \hei\ $\lambda$3889 is fixed to be 88\% of \hei\ $\lambda$5876 theoretically. However, since \hei\ $\lambda$5876 and \hei\ $\lambda$3889 are far separated which makes them suffer from different dust extinctions, we had to correct the flux using the predicted dust extinction value before subtracting the expected \hei\ $\lambda$3889 flux from \hz.

The \hg\ flux detected is pretty consistent with the predictions in all three \nii/\ha\ bins. \hd\ is on the prediction curve in the high \nii/\ha\ bin, but below it in the mid and low \nii/\ha\ bin. \he\ fluxes are above the prediction curves. One of the reasons causing the offsets is that the right sideband (3898.86 - 3912.86 \AA) of \neiii\ $\lambda$3869 is significantly higher than the left sideband due to imperfect stellar continuum subtraction. The fitline then has a positive slope making an underestimation of the flux. As a result, \he\ flux is overmeasured. Since \hz\ shares the similar sidebands with \neiii\ $\lambda$3869, the same effect largely affects the measurement of \hz\ and therefore we do not trust the measurement on the weak \hz\ line which is not included in the figure.

In conclusion, the well-measured high-order Balmer lines are consistent with the prediction of $A_V$ by \ha\ and \hb\ and the \cite{fitzpatrickCorrectingEffectsInterstellar1999} law. In \cite{yanShocksPhotoionizationDirect2018}, higher-order Balmer lines show much less extinction than that derived from \ha/\hb\ and that may be caused by noisier measurement and the difference of \nii/\ha\ binning method. In spite of the consistency between the Balmer lines, the measured degree of dust extinction may not be applicable to the emission lines of other species. This will be discussed in more detail in Sect. \ref{subsec: temperature}. It is possible that different emission lines have different effective extinctions in these low spatial resolution observations, just like what \cite{jiNeedMulticomponentDust2023} found in star-forming galaxies. This finding is consistent with \cite{yanShocksPhotoionizationDirect2018}.

\subsection{Density derivation}
\label{subsec: density}
We derived the electron density using the ratio of \sii\ $\lambda$6731/\sii\ $\lambda$6716. This doublet is optimal for measuring density as it has a very tiny dependence on temperature and dust extinction. We computed the density with the \textsc{PyNeb} package \citep{luridianaPyNebNewTool2015}, which is the Python version of the IRAF package \textsc{NEBULAR}. Assuming the temperature to be 10,000K, the measured electron densities are 45.20$^{+12.32}_{-9.68}$ cm$^{-3}$, 57.52$^{+9.53}_{-12.32}$ cm$^{-3}$, and 42.33$^{+10.37}_{-10.49}$ cm$^{-3}$ for the three \nii/\ha\ bins, respectively. All the measured densities lie in the low-density regime. Collisional de-excitation is not likely to happen in such diffuse regions and it is reliable to measure the temperature using the deduced densities. The measured densities are shown in Table \ref{table: temden}. 

The error of the measured electron density in Table \ref{table: temden} was estimated by repeating the line ratio measurement after adding uncertainties. For pixels in both the sidebands and the central region, we assumed the flux errors follow Gaussian distribution with the width of the Gaussian profile set to be 1$\sigma$ standard deviation propagated from the stacking process. In each measurement, we randomly picked a number in the Gaussian distribution of each pixel as the flux. For each \nii/\ha\ bin, the error was given by the highest and lowest derived density in 10,000 simulations. 

\begin{table*}
    \caption{Density and temperature derived from different line ratios}
    \begin{tabularx}{\linewidth}{@{}l *4{>{\centering\arraybackslash}X}@{}}
    \hline
     Indicators (Unit: cm$^{-3}$ for $n$ and 10$^{4}$ for $T$) & High \nii/\ha   & Mid \nii/\ha   & Low \nii/\ha   \\
     [2pt]
    \hline
    $n_e$ from \sii\ $\lambda$6731/\sii\ $\lambda$6716  & 45.20$^{+12.32}_{-9.68}$    & 57.52$^{+9.53}_{-12.32}$            & 42.33$^{+10.37}_{-10.49}$           \\
    [2pt]
    $T_{\oiii}$ from \oiii\ $\lambda$4363/\oiii\ $\lambda$5007 & 3.17$^{+0.63}_{-0.51}$ & 2.19$^{+0.56}_{-0.48}$ & <1.54            \\
    [2pt]
    $T_{\nii}$ from \nii\ $\lambda$5755/\nii\ $\lambda$6583 & <0.9            & <0.99             & <1.18 \\
    [2pt]
    $T_{\sii}$ from \sii\ $\lambda\lambda$4068, 4076/\sii\ $\lambda\lambda$6716, 6731& 0.94$^{+0.04}_{-0.04}$ & 0.98$^{+0.04}_{-0.04}$   & 1.02$^{+0.06}_{-0.06}$ \\
    [2pt]
    $T_{\oii}$ from \oii\ $\lambda\lambda$7320, 7330/\oii\ $\lambda\lambda$3727, 3729 & 1.38$^{+0.12}_{-0.11}$ & 1.33$^{+0.06}_{-0.05}$   & 1.33$^{+0.07}_{-0.06}$  \\
    [2pt]
    \hline
    \end{tabularx}
    \label{table: temden}
\end{table*}

\begin{table*}
    \caption{Density and temperature derived from different line ratios after correcting for extinction measured by Balmer decrement. These corrections could be significant overcorrections for the low-ionization lines (see  text).}
    \begin{tabularx}{\linewidth}{@{}l *4{>{\centering\arraybackslash}X}@{}}
    \hline
     Indicators (Unit: cm$^{-3}$ for $n$ and 10$^{4}$ for $T$) & High \nii/\ha   & Mid \nii/\ha   & Low \nii/\ha   \\
     [2pt]
    \hline
    $n_e$ from \sii\ $\lambda$6731/\sii\ $\lambda$6716  & 45.20$^{+12.32}_{-9.68}$    & 52.69$^{+14.36}_{-7.49}$            & 40.51$^{+14.54}_{-10.04}$           \\
    [2pt]
    $T_{\oiii}$ from \oiii\ $\lambda$4363/\oiii\ $\lambda$5007 & 3.67$^{+0.33}_{-0.67}$ & 2.49$^{+0.76}_{-0.60}$ & <1.65            \\
    [2pt]
    $T_{\nii}$ from \nii\ $\lambda$5755/\nii\ $\lambda$6583 & <0.94            & <1.05             & <1.25 \\
    [2pt]
    $T_{\sii}$ from \sii\ $\lambda\lambda$4068, 4076/\sii\ $\lambda\lambda$6716, 6731& 1.33$^{+0.08}_{-0.08}$ & 1.66$^{+0.10}_{-0.11}$   & 1.52$^{+0.13}_{-0.13}$ \\
    [2pt]
    $T_{\oii}$ from \oii\ $\lambda\lambda$7320, 7330/\oii\ $\lambda\lambda$3727, 3729 & 0.97$^{+0.06}_{-0.06}$ & 0.87$^{+0.02}_{-0.03}$   & 0.95$^{+0.03}_{-0.04}$  \\
    [2pt]
    \hline
    \end{tabularx}
    \label{table: temden_ext}
\end{table*}

\subsection{Temperature derivation}
\label{subsec: temperature}
For the same ion species, the temperature can be derived from the auroral-to-strong lines ratio. From our emission line measurement, we deduced temperatures of O$^+$, O$^{2+}$, S$^+$, and N$^+$. These ions populate different parts of the interstellar gas in different temperatures. Among the four, \oiii\ lines have the highest ionization potential and therefore represent more ionized regions. The other three ions, O$^+$, S$^+$, and N$^+$, have similar ionization energy, and therefore emission lines from these originate from similar regions with similar temperatures. 

Same as electron densities, temperatures were calculated using \textsc{PyNeb}. For undetected lines, the upper limit of temperature is given by assuming the flux value of the line to be 2$\sigma$. The upper and lower bounds of the measurement were calculated by adding 1$\sigma$ uncertainty derived from the sliding box as described in Sect. \ref{subsec: measurement}. The uncertainty in electron density was also considered. Using the fact that the effect of electron density on temperature measurement is monotonic, we computed the temperatures using the maximum and minimum electron density given in Sect. \ref{subsec: density}.

All the temperatures derived without and with extinction correction are given in Table \ref{table: temden} and Table \ref{table: temden_ext}, respectively. However, we believe that those without extinction correction are closer to the truth than those corrected using the extinction measured by Balmer decrement. The reason is explained below. Here we quote the temperatures using those in Table \ref{table: temden}, without extinction correction. We discuss the results of different ions following the order of high, mid, and low \nii/\ha\ bins:
\begin{enumerate}
    \item O$^{2+}$ zone: \oiii\ $\lambda$4363 was only detected in the high and mid \nii/\ha\ bins as the flux of \oiii\ $\lambda$4363 significantly decreases with metallicity and becomes undetectable in the low \nii/\ha\ bin. We note that some of the flux from the measured \oiii\ $\lambda$4363 flux may be contributed by \feii\ $\lambda$4359, which can be estimated by another forbidden line \feii\ $\lambda$4287. However, no visible bump was found around 4288\AA\ in all stacked spectra. Therefore, we did not attempt any correction on the strength of \oiii\ $\lambda$4363. The temperatures of the O$^{2+}$ zone in the three \nii/\ha\ bins were measured to be 3.17$^{+0.63}_{-0.51}$ $\times$ 10$^4$K, 2.19$^{+0.56}_{-0.48}$ $\times$ 10$^4$K and < 1.54 $\times$ 10$^4$K. Consistent with the findings of \cite{yanShocksPhotoionizationDirect2018}, the temperature of the O$^{2+}$ zone increases with metallicity, contrary to the common wisdom. With a higher abundance of metal, the ionized gas should be more efficiently cooled, resulting in a lower temperature. This measurement raises the question of whether metallicity is affecting ionic temperature through other mechanisms in LIERs, or whether our binning of \nii/\ha\ might be tracing other properties.\\
    \item N$^+$ zone: Since \nii\ $\lambda$5755 was not detected in all \nii/\ha\ bins, we gave an upper limit for its temperature. Using 2$\sigma$ value as the flux of \nii\ $\lambda$5755, the derived temperature limits are < 0.9 $\times$ 10$^4$K, < 0.99 $\times$ 10$^4$K and < 1.18 $\times$ 10$^4$K in the three bins, respectively.\\
    \item S$^+$ zone: We had very robust detections on \sii\ $\lambda\lambda$4068, 4076. The temperatures derived are 0.94$^{+0.04}_{-0.04}$ $\times$ 10$^4$K, 0.98$^{+0.04}_{-0.04}$ $\times$ 10$^4$K and 1.02$^{+0.06}_{-0.06}$ $\times$ 10$^4$K. The S$^+$ zone temperatures do not show large variations among the three bins. It is not surprising that the temperature decreases when metallicity increases. Given that neutral nitrogen atoms have similar ionization energy to neutral sulfur atoms, the temperatures of the N$^+$ zones may be also close to the S$^+$ zones.\\
    \item O$^+$ zone: \oii\ $\lambda\lambda$7320, 7330 were significantly detected in all three bins. \oii\ $\lambda\lambda$7320, 7330 overlaps with another emission line \caii\ $\lambda$7324 which can be estimated through \caii\ $\lambda$7291. Nevertheless, \caii\ $\lambda$7291 was not detected in all stacked spectra. Therefore, we assumed the contribution of \caii\ $\lambda$7324 to the \oii\ auroral lines is negligible. From the measurement, we derived the temperature of O$^+$ zone to be 1.38$^{+0.12}_{-0.11}$ $\times$ 10$^4$K, 1.33$^{+0.06}_{-0.05}$ $\times$ 10$^4$K and 1.33$^{+0.07}_{-0.06}$ $\times$ 10$^4$K in the three bins. This is consistent with the expectation that the O$^+$ zone temperatures are higher than the S$^+$ zone because neutral oxygen atoms have a higher ionization energy than neutral sulfur atoms.
\end{enumerate}

The reason that we prefer the temperatures derived without extinction correction is that applying the dust extinction derived in Sect. \ref{subsec: balmer} will lead to an unphysical scenario similar to that encountered by \cite{yanShocksPhotoionizationDirect2018}. After correction, the temperatures of S$^+$ zone would be much higher than the O$^+$ zone, by 37\% to 90\% in different bins. Based on the first ionization energies of neutral oxygen atoms and neutral sulfur atoms, the S$^+$ zone spans deeper toward the neutral gas compared to the O$^+$ zone. The line in Fig. \ref{fig:S2O2} shows the \sii\ and \oii\ line ratios when the temperatures of the two zones are the same. The shock models predict S$^+$ zone to have a lower temperature due to the reason stated above. The photoionization model grid is located slightly above the line which means that the S$^+$ zone has slightly higher temperatures than the O$^+$ zone because fewer coolants are available in deeper regions and ionizing radiation becomes harder. Nevertheless, none of the models predict a temperature of S$^+$ zone to be a lot higher than the O$^+$ zone. Therefore the extinction measured from the Balmer decrement is likely inapplicable for the correction of O$^+$ and S$^+$ lines. To get rid of the effect of dust, we constructed a dust-insensitive temperature indicator combining the auroral-to-strong line ratios of \sii\ and \oii. This will be discussed in Sect. \ref{sec: Discussion}.
\section{Discussion}
\label{sec: Discussion}
In this section, we compare the measured temperatures to the predictions of models by the two possible ionization mechanisms of LIERs, photoionization from hot, evolved stars and fast radiative shocks. In Sect. \ref{subsec: Models} we describe the models we are using in this study. The measured temperature-sensitive line ratios will be compared to the theoretical values predicted by the models in Sect. \ref{subsec: Compare}.

\subsection{Shocks and photoionization models}
\label{subsec: Models}
\begin{enumerate}
    \item The photoionization models: The models were computed with CLOUDY c17.02 \citep{ferland2017ReleaseCloudy2017}. The ionizing SED was generated by Python-FSPS (Flexible Stellar Population Synthesis)\citep{conroyPropagationUncertaintiesStellar2009, conroyPropagationUncertaintiesStellar2010}, assuming a simple stellar population (SSP) from a starburst of 5 Gyr ago. At this age, ionizing photons are primarily contributed by post-asymptotic giant branch (pAGB) stars. We used a Kroupa IMF \citep{kroupaVariationInitialMass2001, kroupaGalacticFieldInitialMass2003} and MIST isochrones \citep{choiMesaIsochronesStellar2016, dotterMESAIsochronesStellar2016} for the SSP. Each set of SSPs contains 6 different stellar metallicities: log(Z/Z$_\odot$) = -0.6, -0.4, -0.2, 0.0, 0.2, 0.4. The solar abundance set we used is taken from \cite{grevesseChemicalCompositionSun2010}, where 12 + log(O/H) = 8.69. We matched the gas-phase metallicities with the stellar metallicities when computing photoionization models. In reality, the stellar metallicity does not need to be close to the gas-phase metallicity in these early-type galaxies. Regardless, the dependence of the shape of the ionizing spectrum on the stellar metallicity is small if pAGB stars are the dominant population \citep{bylerSelfconsistentPredictionsLIERlike2019}. For each photoionization model, there are 4 different ionization parameters: log(U) = -4.5, -3.5, -2.5, -1.5. The ionization parameter U is defined as the ionizing flux divided by the gas density. For the elements having secondary origins including carbon and nitrogen, we assumed their relative abundances to oxygen change with O/H following the prescriptions given by \cite{grovesDustyRadiationPressureDominated2004}. We chose not to use the relation given by \cite{dopitaNewStronglineAbundance2013} as such photoionization models are not able to cover the data points in Fig. \ref{fig:metal}. This is consistent with \cite{jiConstrainingPhotoionizationModels2020} that only the prescription of \cite{grovesDustyRadiationPressureDominated2004} is able to cover the whole AGN/LIER region in the \nii/\ha\ BPT diagram.  Finally, we used the default dust depletion factors of CLOUDY v17.02 to remove part of the elements in the gas phase that condense onto dust grains \citep{cowieHighresolutionOpticalUltraviolet1986, jenkinsElementAbundancesInterstellar1987}.\\
    \item The shock models: We used the shock models from the 3MdBs database by \cite{alarieExtensiveOnlineShock2019}, which contains models generated using version 5.1.13 of \textsc{MAPPINGS V} \citep{sutherlandMAPPINGSAstrophysicalPlasma2018, sutherlandEffectsPreionizationRadiative2017}. The fast radiative shock models in the database cover a large range of pre-shock densities $n$, magnetic fields $B$ and shock velocities $v$. Using different combinations of $B$ and $n$, different values of magnetic field parameter ($B/n^{1/2}$) can be covered. We only included shocks without precursors grid since the amount of gas beyond the shock is insufficient to form an ionized precursor in LIERs \citep{dopitaSpectralSignaturesFast1995}. The typical compression factor after the shock passed through the gas is 4 and the magnetic field might reduce the number by providing pressure support. When the gas started to cool, the gas was further compressed and the density increases. It is hard to calculate the exact compression factor because it depends on the magnetic field, the shock velocity, the pre-shock density, and the temperature. However, in the shock models we consider, the compression factor is between 100 and 1000 after the cooling phase is finished. Compared to our measured density, which is about 50 cm$^{-3}$, the model grids of n = 1 cm$^{-3}$ and n = 10 cm$^{-3}$ might be able to represent the gas in different stages of cooling and we computed the ratio of \sii\ doublets for checking. We included both of the model grids with solar metallicities. Variation in metallicities may also affect the line ratio, and therefore we also plotted a grid with sub-solar LMC metallicity and n = 1 cm$^{-3}$ for reference. For each of these grids, there are 4 different shock velocities: V = 200, 300, 400 and 500 km s$^{-1}$, and 5 magnetic parameters: B/n$^{1/2}$ = 0, 1, 2, 4, 5 $\mu$G cm$^{3/2}$. Since there are no B/n$^{1/2}$ = 0 $\mu$G cm$^{3/2}$ models, we chose the weakest magnetic field model available. For the two n = 1 cm$^{-3}$ models, the weakest magnetic field available gives B/n$^{1/2}$ = 10$^{-4}$ $\mu$G cm$^{3/2}$. For the n = 10 cm$^{-3}$ model, B/n$^{1/2}$ = 3.16 $\times$ 10$^{-4}$ $\mu$G cm$^{3/2}$ is the weakest option. Dust depletion is not considered in the shock models since fast shocks will destroy dust by grain-grain collisions \citep{allenMAPPINGSIIILibrary2008}. Even if there is remaining dust depletion, our conclusion will not be largely affected since we are looking at ratios of lines from the same ions, except \nii/\ha\ in Fig. \ref{fig:SOTN2Ha}. To check the presence of dust depletion, we compare the flux of \caii\ $\lambda$7291 with the predictions from the shock models. The shock models indicate the strength of \caii\ $\lambda$7291 should be at least 10\% to 50\% of \hb\ depending on the shock velocities, magnetic parameter and metallicity. Nevertheless, we did not detect the line in our spectrum. The 2$\sigma$ upper limits are  9.5\%, 7.5\%, and 3.8\% of their respective \hb\ flux in the high, mid, and low \nii/\ha\ bins. With the fact that the strength of \caii\ $\lambda$7291 is predicted to be stronger in lower metallicity, we concluded that dust depletion is present in the gas.
\end{enumerate}

\subsection{Data-model comparison}
\label{subsec: Compare}
\begin{figure}
        \includegraphics[width=\columnwidth]{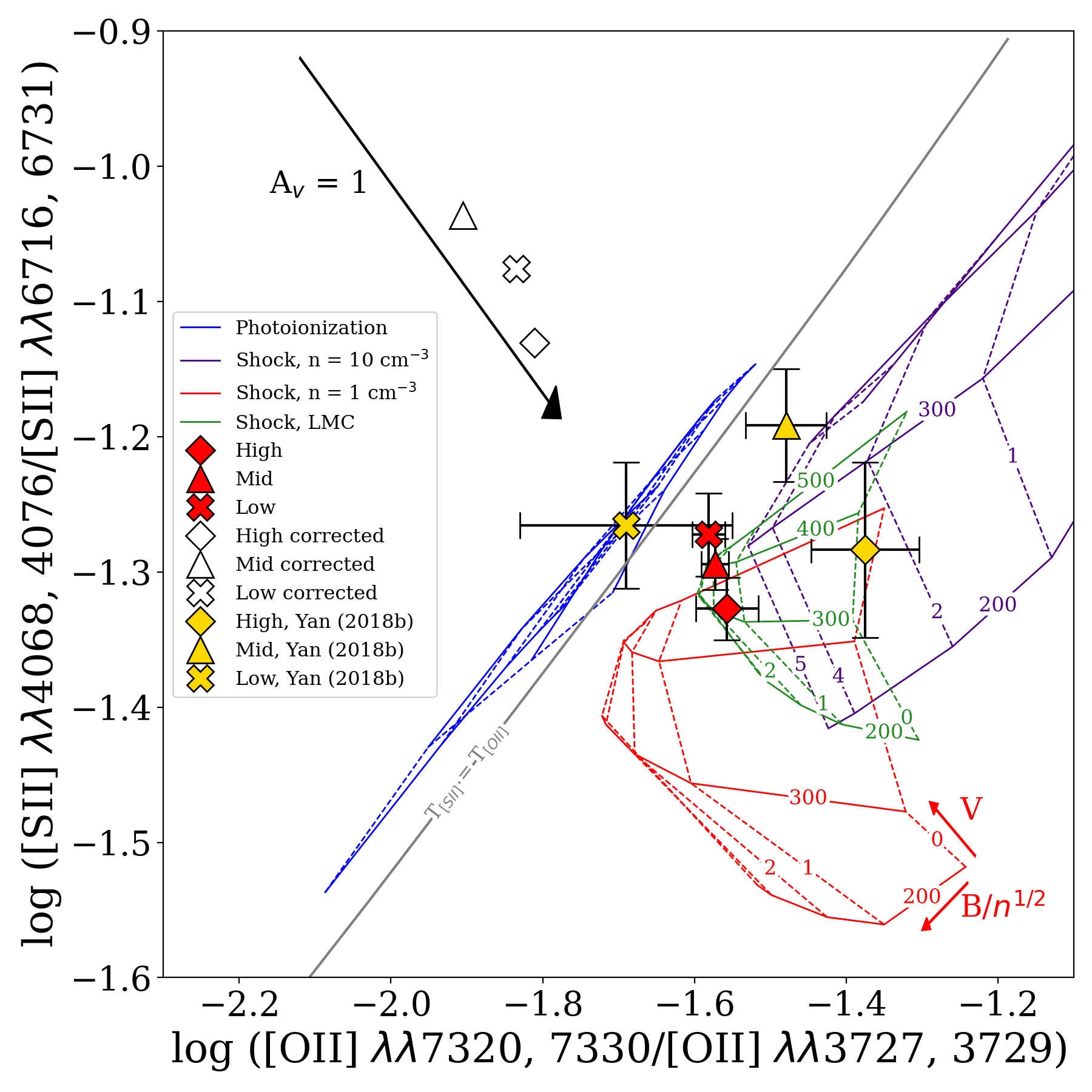}
    \caption{log (\sii\ $\lambda\lambda$4068, 4076/ \sii\ $\lambda\lambda$6716, 6731) vs. log (\oii\ $\lambda\lambda$7320, 7330/ \oii\ $\lambda\lambda$3727, 3729). All the data points are plotted with error bars indicating 1$\sigma$ uncertainty. The blue grid is the photoionization model and the other three grids are the shock models. The indigo grid has a pre-shock density of n = 10 cm$^{-3}$ and solar metallicity. The red grid has a pre-shock density of n = 1 cm$^{-3}$ and solar metallicity. The green grid has a pre-shock density of n = 1 cm$^{-3}$ and sub-solar LMC metallicity. The black arrow represents the effect of 1 mag extinction in $A_V$. Extinction correction using the coefficient predicted by the Balmer decrement will move the data points to the locations indicated by the white data points, which cannot be explained by any model and current understanding of ISM. Therefore, the correction may not be physical. The gray line represents the line ratios when the temperature measured by \sii\ is the same as \oii.}
    \label{fig:S2O2}
\end{figure}

\begin{figure}
        \includegraphics[width=\columnwidth]{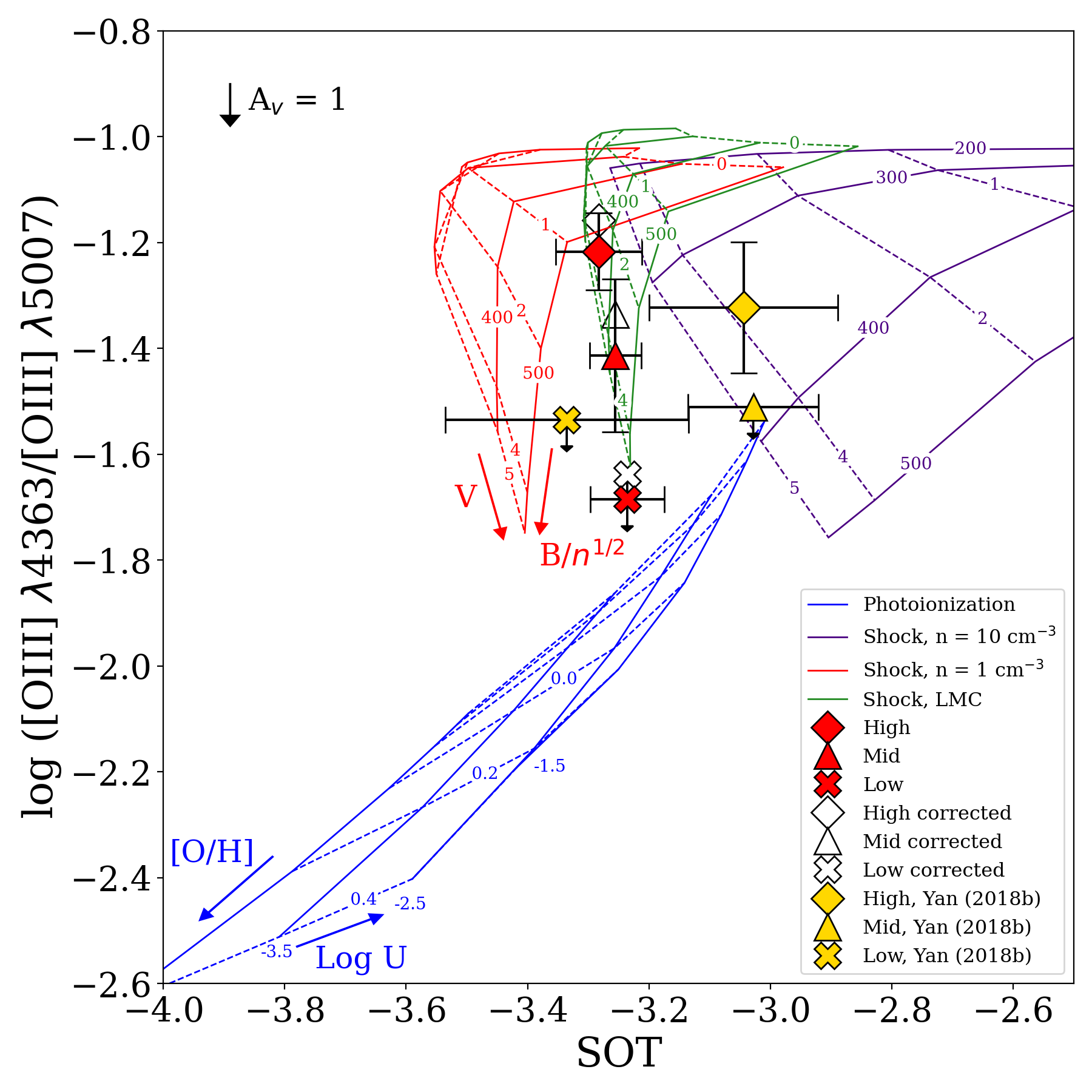}
    \caption{log (\oiii\ $\lambda$4363/ \oiii\ $\lambda$5007) vs. SOT. The error bars, the grids, and the black arrow are the same as those in Fig. \ref{fig:S2O2}. The dust extinction corrected data points are in white. These corrections could be significant overcorrections for the low-ionization lines (see text). With a given value of SOT, the photoionization model prefers a lower O$^{2+}$ zone temperature than the shock models. For undetected lines, the upper limits are denoted by arrows and are calculated using the 2$\sigma$ flux values.}
    \label{fig:O3SOT}
\end{figure}

\begin{figure}
        \includegraphics[width=\columnwidth]{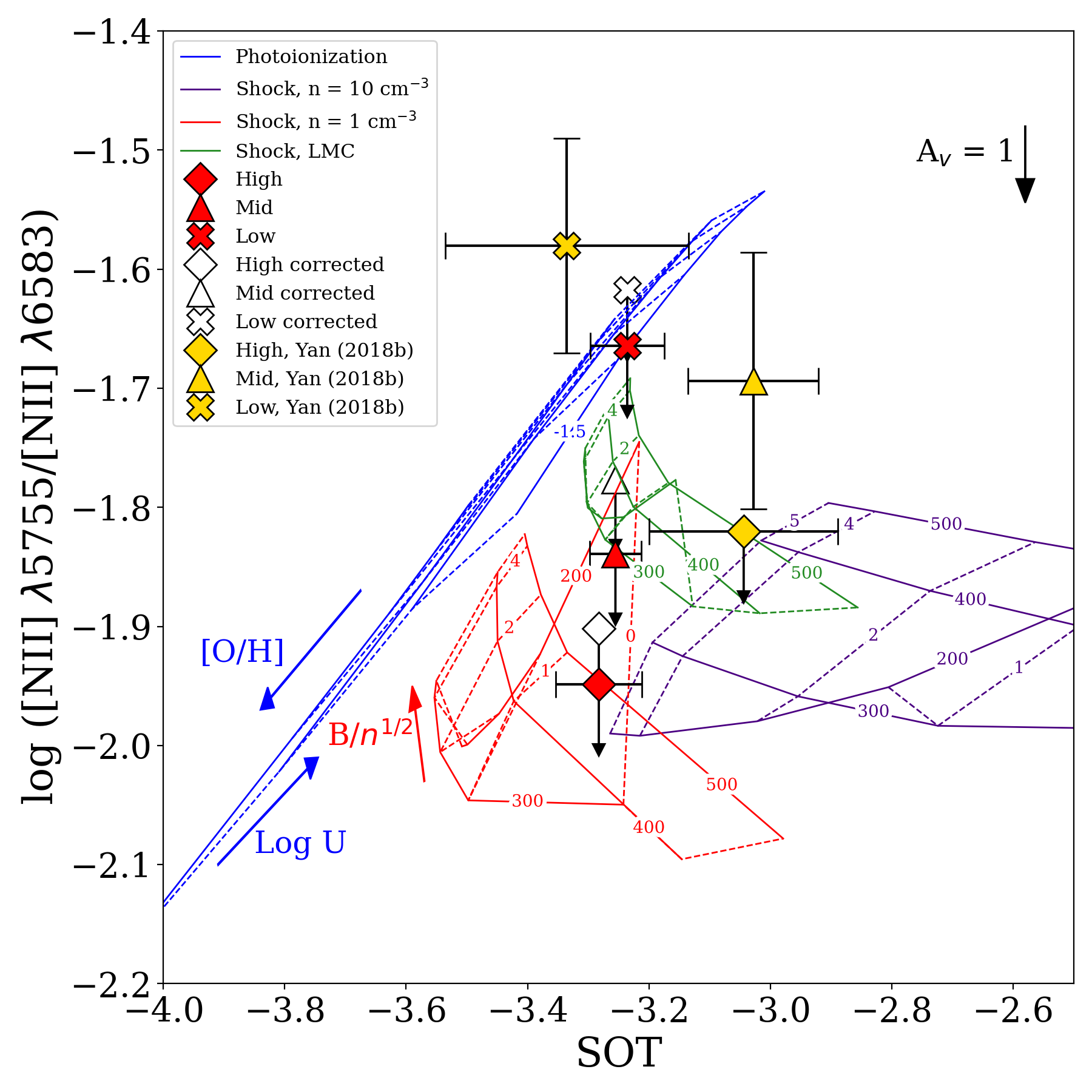}
    \caption{log (\nii\ $\lambda$5755/ \nii\ $\lambda$6583) vs. SOT. The error bars, the grids, and the black arrow are the same as those in Fig. \ref{fig:S2O2}. The dust extinction corrected data points are in white. These corrections could be significant overcorrections for the low-ionization lines (see text). For undetected lines, the upper limits are denoted by arrows and are calculated using the 2$\sigma$ flux values.}
    \label{fig:N2SOT}
\end{figure}

\begin{figure}
        \includegraphics[width=\columnwidth]{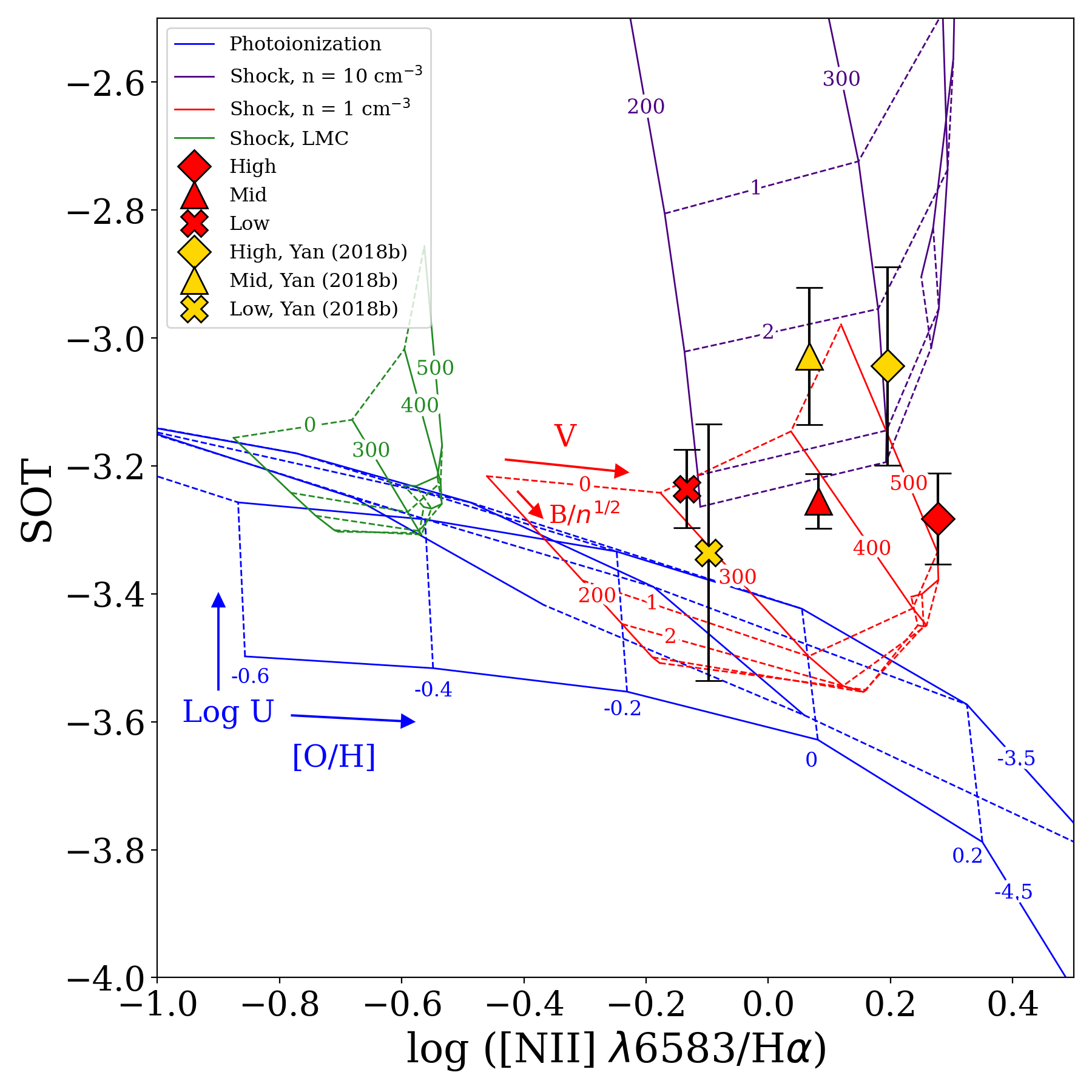}
    \caption{log (\nii\ $\lambda$6583/ \ha) vs. SOT. Both values are insensitive to dust, and therefore the extinction vector is not drawn. The error bars on the x-axis are also neglected as they are smaller than the marker and the error bars indicate 1$\sigma$ uncertainties. All the grids are the same as those in Fig. \ref{fig:S2O2}. The blue photoionization grid predicts a lower SOT value compared to all of the data points. All of the data support the interstellar shock models. However, the actual shock parameters cannot be tightly constrained due to the uncertainty in metallicity and pre-shock density.}
    \label{fig:SOTN2Ha}
\end{figure}

In Figs. \ref{fig:S2O2} to \ref{fig:SOTN2Ha}, we included the model grids mentioned in the last subsection, the measured line ratios from this study and those from \cite{yanShocksPhotoionizationDirect2018}. We note that the data from \cite{yanShocksPhotoionizationDirect2018} is based on the SDSS Legacy. They are not observed using IFUs, but are done using single fiber centered on the centers of galaxies. The model used in this study is also not the same as that used by \cite{yanShocksPhotoionizationDirect2018}. In \cite{yanShocksPhotoionizationDirect2018}, the shock models applied are from \textsc{MAPPINGS III} and in this paper, models from \textsc{MAPPINGS V} are used. Two main differences exist between the photoionization models in this paper and those used by \cite{yanShocksPhotoionizationDirect2018}. First, the ionizing spectra are slightly different. In \cite{yanShocksPhotoionizationDirect2018}, the photoionization model was generated using a 13-Gyr-old SSP from \cite{bruzualStellarPopulationSynthesis2003}. In such a model, the isochrone is computed by the Padova 1994 stellar evolutionary tracks \citep{alongiEvolutionarySequencesStellar1993, bressanEvolutionarySequencesStellar1993, fagottoEvolutionarySequencesStellar1994, fagottoEvolutionarySequencesStellar1994a, girardiEvolutionarySequencesStellar1996}, the STELIB/BaSeL 3.1 spectral library, and the Chabrier Initial Mass Function. The STELIB/BaSeL 3.1 uses STELIB spectra \citep{leborgneSTELIBLibraryStellar2003}, which cover the wavelength range of 3200-9500 \AA\ and are extended into infrared and ultraviolet using the color--temperature calibrations of BaSeL 3.1 `WLBC 99' SED library \citep{westeraStandardStellarLibrary2002}. In this study, we assume a 5-Gyr-old SSP from FSPS using the MIST isochrone \citep{choiMesaIsochronesStellar2016, dotterMESAIsochronesStellar2016}. Second, the N/O ratio in \cite{yanShocksPhotoionizationDirect2018} follows the relation from \cite{vila-costasNitrogentooxygenRatioGalaxies1993}, and this study uses the relations in \cite{grovesDustyRadiationPressureDominated2004} instead, which better describes the AGN/LIER spaxels in MaNGA \citep{jiCorrelationGasphaseMetallicity2022}.

Figure \ref{fig:S2O2} shows the temperature-sensitive line ratios for \sii\ and \oii. As mentioned in Sect. \ref{subsec: temperature}, the partially ionized zone of singly ionized sulfur extends deeper toward the neutral zone and should have a similar, or lower temperature than the O$^+$ zone. This is predicted by both the shocks and photoionization models, which only cover the central and the lower right regions of the figure. The extinction coefficient predicted by the Balmer decrement is around A$_V\sim$0.7-0.9. If we apply extinction correction using these A$_V$ values, all data points measured would shift to the `unphysical' upper left region, as shown by the data points in white. The gray line shows the line ratios where the temperature measured by \sii\ is the same as \oii. We computed the line by \textsc{PYNEB} assuming the density to be 50 cm$^{-3}$. Varying the density within 1-100 cm$^{-3}$ will induce a small shift to the line and will not change the conclusion.

All of our data are located near the n = 1 cm$^{-3}$ shock model with a shock velocity of around 500 km s$^{-1}$ and magnetic parameter between 0 and 1 $\mu$G cm$^{3/2}$. Looking at the position of other shock model grids, both an increase in pre-shock density and a decrease in metallicity will cause the models to move upward. Therefore, it is also possible that our measurements do not indicate such high-velocity shocks, but an environment with sub-solar metallicity or higher density, or a combination of both. Our data are generally consistent with the increasing S$^+$ temperature with decreasing metallicity of shock models. Looking at the SDSS-I data, the high \nii/\ha\ data is on the n = 10 cm $^{-3}$ grid with a pre-shock velocity between 200 to 300 km s$^{-1}$ and magnetic parameter between 2 and 3 $\mu$G cm$^{3/2}$. It is reasonable as the post-shock density measured for that bin was 136 $\pm 14$ cm$^{-3}$, which is significantly higher than any other bins. The mid \nii/\ha\ SDSS-I data is not on any of the grids. That set of data has lower metallicity and density compared to the high \nii/\ha\ data. While a decrease in metallicity may cause the point to move upward, it is not expected for these galaxies to have metallicity lower than the LMC. In contrast, the low \nii/\ha\ Sloan data is consistent with the photoionization model. It is important to note that we do not account for any dust extinction effect in the above discussion. Since it is expected that the line ratios are affected by a certain degree of dust extinction, we cannot rule out the photoionization model from this plot alone.

While both \sii\ and \oii\ line ratios are sensitive to dust, a dust-insensitive temperature indicator, SOT, can be constructed using the combination of the two line ratios. This was proposed by \cite{yanNitrogentoOxygenAbundanceRatio2018}. Since the auroral \oii\ lines are on the red side and its strong lines are on the blue side, introducing dust extinction will increase the ratio. On the contrary, the \sii\ auroral lines are on the blue side and strong lines are on the red side. As a result, we can combine the two line ratios to make a dust-insensitive temperature indicator. The definition of SOT in this paper is 
\begin{equation}
    \centering
    \text{SOT} = \log\frac{\oii\ \lambda\lambda \text{7320, 7330}}{\oii\ \lambda\lambda \text{3727, 3729}} + \text{1.3} \times \log\frac{\sii\ \lambda\lambda \text{4068, 4076}}{\sii\ \lambda\lambda \text{6716, 6731}}.
    \label{eq:SOT}
\end{equation}
We note that the definition is slightly different from the original paper. We put the weak lines in the numerator to be consistent with the signs in other plots. The constant 1.3 is to account for the difference in wavelength separation and has negligible dependence on the choice of extinction curve.

A plot of log (\oiii $\lambda$4363/\oiii $\lambda$5007) versus SOT is given in Fig. \ref{fig:O3SOT}. From the extinction vector, it can be seen that SOT is insensitive to dust, any dust extinction will only move the data points in the vertical direction. The high and mid-\nii/\ha\ data are consistent with the LMC shock models with magnetic parameters between 2 and 4 $\mu$G cm$^{3/2}$. The high-metallicity data has a pre-shock velocity of 300 to 400 km s$^{-1}$ and the mid metallicity data has one between 400 to 500 km s$^{-1}$. However, quiescent galaxies are not likely to have metallicities similar to the LMC. We note that SOT shows a degeneracy between pre-shock density and metallicity. Thus, we suspect the positions of the data could be consistent with a pre-shock density higher than 1 cm$^{-3}$ and a metallicity close to solar value. The SOT is sensitive to both metallicity and pre-shock density, and the actual pre-shock density values of these data might be between 1 cm$^{-3}$ and 10 cm$^{-3}$ which cause the data points to fall between the two models. The data of these two bins provide strong evidence for the interstellar shocks because the photoionization model cannot explain such high temperature in the O$^{2+}$ zone. The \oiii\ measurement of low metallicity gives a 2$\sigma$ upper limit. It is hard to distinguish whether it is consistent with the shock, or photoionization model.

Another point to note is how the locations of the data vary with metallicity. Both the shock models and photoionization models predict the low \nii/\ha\ bin will be shifted to the upper right with respect to higher \nii/\ha\ bins. But our data show the opposite trend. None of the models can explain the positive correlation between \oiii\ temperature and metallicity as we mentioned in Sect. \ref{subsec: temperature}. On the other hand, for the shock models, increasing the magnetic parameters and shock velocities matches the tendency of the variation. It is possible that these variables might be correlated with the metallicity in the shock mechanism, or related with the increase of \nii/\ha\ ratio.

Considering the SDSS-I data points, the high-metallicity data lies on the 10 cm$^{-3}$ data grid which has solar metallicity. It has a $B/n^{1/2} \approx 4$ $\mu$G cm$^{3/2}$ and pre-shock velocity between 300 and 400 km s$^{-1}$. \oiii\ $\lambda$4363 is not detected in mid and low \nii/\ha\ bins. Interestingly, the low-metallicity data is on the left of the other two bins. This contradicts what both the shock and photoionization models predict: when metallicity decreases, the data will go to the right. Nevertheless, this is suspected to be also the effect of pre-shock density. In the measurement of \cite{yanShocksPhotoionizationDirect2018}, the post-shock density of the low \nii/\ha\ bin is significantly lower than the remaining bins, which is 34 $\pm$ 15 cm$^{-3}$ compared to 136 $\pm$ 14 cm$^{-3}$ and 71 $\pm$ 12 cm$^{-3}$ for the high and mid bins, respectively. As a result, the position of the low \nii/\ha\ bin data may be affected by density.

Figure \ref{fig:N2SOT} shows the data in log (\nii\ $\lambda$5755/\nii\ $\lambda$6583) versus SOT plot. Although none of our data can detect \nii\ $\lambda$5755, the upper limits of the three bins all show that the photoionization model predicts a higher \nii\ temperature than the measured values. The trend of the upper limit shows a similar relationship to the model predictions. Compared to our data, the SDSS-I data show a larger variation. Since the x-axis is the same as Fig. \ref{fig:O3SOT}, we believe the horizontal position difference between the low \nii/\ha\ bin data and the rest of the bins is caused by the difference in pre-shock density. While the high \nii/\ha\ data favors the shock models, SDSS-I data of the mid and low \nii/\ha\ bins are not covered by any of the grids. It is possible that the position of SDSS-I mid-metallicity data contains both the effects of pre-shock density and metallicity. We do not have a model grid with n = 10 cm$^{-3}$ and sub-solar metallicity which might be able to cover the data point. However, it is important to note again that the SDSS quiescent galaxies are not likely to have a metallicity as low as the LMC. Last but not least, the SDSS-I low \nii/\ha\ data shows a higher \nii\ ratio than all model predictions and the reason remains unknown.

Finally, SOT versus log(\nii\ $\lambda$6583/\ha) is given in Fig. \ref{fig:SOTN2Ha}. Both of the line ratios are dust insensitive and the dust extinction vector is too small to be shown. Since both \nii\ $\lambda$6583 and \ha\ are robustly detected, we did not include the error bars in the x direction which are smaller than the marker. This figure provides a shred of evidence to rule out the photoionization model which predicts a smaller SOT than the measured values. Our measured data are consistent with the solar metallicity shock model with n = 1 cm$^{-3}$. The high, mid, and low \nii/\ha\ points have $B/n^{1/2}$ between 0 to 1 $\mu$G cm$^{3/2}$ and pre-shock velocity of 500, 400, and 300 km s$^{-1}$ respectively. However, the position of the data points on the grid depends not only on these two parameters but also on metallicity. By comparing the LMC metallicity grid and the solar metallicity grid, a smaller metallicity will shift the model to the left as predicted by the \nii/\ha. It is worth mentioning that from the model grids in the figure, \nii/\ha\ does not solely trace metallicity. In shock models, \nii/\ha\ increases with shock velocity while in photoionization models, it is sensitive to the ionization parameter. Therefore, the binning of \nii/\ha\ might not efficiently divide the data based on metallicity. Without accurate measurement of metallicity, we cannot constrain the model grid to be used, and the actual shock velocity and magnetic parameter cannot be tightly constrained.

From the shock models, SOT is not only sensitive to temperature but also the pre-shock density. Again, the difference in the SOT for the SDSS-I data should be attributed to density variations. The high and mid metallicity SDSS-I data are on the n = 10 cm$^{-3}$ shock model. The high \nii/\ha\ Sloan data has a magnetic parameter between 2 to 4 $\mu$G cm$^{3/2}$ and a shock velocity of 300 km s$^{-1}$. The mid \nii/\ha\ data has a similar magnetic parameter and a slightly lower shock velocity. However, it may also be the effect of metallicity difference and it is not possible to constrain shock parameters of the data without a more precise metallicity and pre-shock density value. The low-metallicity SDSS-I data point shows consistency with the n = 1 solar metallicity model, having a shock velocity of 300 km s$^{-1}$ and a magnetic parameter between 0 and 1 $\mu$G cm$^{3/2}$. We emphasize that this data point has the lowest post-shock density measured, which can lower the measured SOT value. Overall, all of the data points favor the shock models.

\section{Conclusions}
\label{sec: conclusions}
In this paper we constrained the ionization source of LIERs by measuring the temperatures of different ions inside these regions in red non-star-forming galaxies. We selected quiescent galaxies from the spatially resolved MaNGA survey and further selected the spaxels that contain LIERs. We divided the sample into three bins based on \nii/\ha\ and stacked them together. For each \nii/Ha bin, we constructed a stellar continuum based on the gas-poor spaxels. The stellar continuum was then subtracted from the stacked spectra of spaxels with strong gas emissions to obtain spectra with purely gas emission lines. Finally, we measured the temperatures using the auroral-to-strong line flux ratios of different ions and compared them to the models of photoionization and interstellar shocks. A detailed summary of our results follows:
\begin{enumerate}
    \item The Balmer decrement overestimated the dust extinction effect experienced by low-ionization ion species. Since the first ionization energy of neutral sulfur is lower than neutral oxygen, we expect the S$^{+}$ zone to have a temperature that is similar to or lower than the O$^{+}$ zone. Nevertheless, from the measurements of \sii\ $\lambda$4068, 4076/\sii\ $\lambda$6716, 6731,  and \oii\ $\lambda$7320, 7330/\oii\ $\lambda$3727, 3729, correcting the extinction effect with the extinction coefficient predicted by the Balmer line ratio will lead to a S$^{+}$ zone that is much hotter than the O$^{+}$ zone, which is not explained by any of the models and physical pictures. As a result, we suspect that the dust extinction derived from the Balmer lines should not be applied to other emission lines.\\
    \item The O$^{2+}$ zone temperature increases with metallicity, which contradicts the general expectation. In general, metals are effective coolants and temperatures are expected to have an inversely proportional relationship with metallicity. In this study we divided the sample into three bins based on \nii/\ha, which is a dust-insensitive metallicity indicator when other parameters are fixed. The temperature derived from \oiii\ $\lambda$4363/\oiii\ $\lambda$5007 in the high, mid, and low \nii/\ha\ bins are found to be 3.17$^{+0.63}_{-0.51}$ $\times$ 10$^{4}$K, 2.19$^{+0.56}_{-0.48}$ $\times$ 10$^{4}$K, and <1.54 $\times$ 10$^{4}$K, respectively. Therefore, there could be other unconsidered processes affecting the O$^{2+}$ zone temperature.\\
    \item The measured data are mostly consistent with the interstellar shock models. Most of the derived line ratios are predicted by the interstellar shock model with n = 1 cm $^{-3}$ and solar metallicity. On the other hand, the high and mid \nii/\ha\ data from \cite{yanShocksPhotoionizationDirect2018} are matched by the shock model with n = 10 cm $^{-3}$ in some of the cases. The intrinsic shock parameters cannot be tightly constrained since the predicted line ratios depend not only on shock velocity and magnetic parameters, but also on pre-shock density and metallicity. Uncertainties in pre-shock density and metallicity estimation make it more difficult to determine the shock parameters.
\end{enumerate}
We note here that our conclusion is different from that reached by \cite{yanShocksPhotoionizationDirect2018}. \cite{yanShocksPhotoionizationDirect2018} concluded that neither photoionization nor shock could explain the measured auroral-to-strong line ratios. That conclusion was based on an older shock model. With the new shock model adopted in this paper, the \cite{yanShocksPhotoionizationDirect2018} measurements are generally consistent with shock models as well.

This work provides strong evidence for interstellar shocks as the ionization source of LIERs. It also shows the power of spatially resolved IFU data. Using such data, smaller-scale structures inside a galaxy can be investigated, giving conclusions different from previous investigations. In the future, continuous development of integral field spectroscopy with deeper observations will help detect weak auroral lines in LIERs. Such measurements can provide even more robust constraints on the current mystery. 

\begin{acknowledgements}
M.-Y.L.L. and R.Y. acknowledge the support of two grants from the Research Grants Council of the Hong Kong Special Administrative Region, China [Project No: CUHK 14303123, CUHK 14302522, ] and the Direct Grant of CUHK Faculty of Science. R.Y. also acknowledges support from the Hong Kong Global STEM Scholar scheme, by the Hong Kong Jockey Club Charities Trust through the JC STEM Lab of Astronomical Instrumentation. Z.S.L. acknowledges the support from Hong Kong Innovation and Technology Fund through the Research Talent Hub program (GSP028). M.-Y.L.L thanks Ricky Wai Kiu Wong and Ziming Peng for their helpful discussions. 

This research makes use of data from the SDSS-IV. Funding for the Sloan Digital Sky Survey IV has been provided by the Alfred P. Sloan Foundation, the U.S. Department of Energy Office of Science, and the Participating Institutions. SDSS acknowledges support and resources from the Center for High-Performance Computing at the University of Utah. The SDSS web site is www.sdss.org. SDSS is managed by the Astrophysical Research Consortium for the Participating Institutions of the SDSS Collaboration including the Brazilian Participation Group, the Carnegie Institution for Science, Carnegie Mellon University, the Chilean Participation Group, the French Participation Group, Harvard-Smithsonian Center for Astrophysics, Instituto de Astrofísica de Canarias, The Johns Hopkins University, Kavli Institute for the Physics and Mathematics of the Universe (IPMU) / University of Tokyo, the Korean Participation Group, Lawrence Berkeley National Laboratory, Leibniz Institut für Astrophysik Potsdam (AIP), Max-Planck-Institut für Astronomie (MPIA Heidelberg), Max-Planck-Institut für Astrophysik (MPA Garching), Max-Planck-Institut für Extraterrestrische Physik (MPE), National Astronomical Observatories of China, New Mexico State University, New York University, University of Notre Dame, Observatório Nacional / MCTI, The Ohio State University, Pennsylvania State University, Shanghai Astronomical Observatory, United Kingdom Participation Group, Universidad Nacional Autónoma de México, University of Arizona, University of Colorado Boulder, University of Oxford, University of Portsmouth, University of Utah, University of Virginia, University of Washington, University of Wisconsin, Vanderbilt University, and Yale University.
\end{acknowledgements}



\bibliographystyle{aa} 

\end{document}